\documentclass[a4paper,11pt]{article}

\usepackage{jheppub}

\usepackage[utf8]{inputenc}
\usepackage[english]{babel}

\usepackage{amsfonts}
\usepackage{amsmath}
\usepackage{amssymb}
\usepackage{mathrsfs}
\usepackage{slashed}
\usepackage{bm}

\usepackage{graphicx}
\usepackage{epsfig}
\usepackage{xcolor}

\usepackage{booktabs}
\usepackage{multirow}
\usepackage{subfig}
\usepackage{caption}
\usepackage{verbatim}
\usepackage[normalem]{ulem}
\usepackage{afterpage}
\usepackage{comment}
\usepackage{hyperref}
\newenvironment{Eqnarray}{\arraycolsep 0.14em\begin{eqnarray}}{\end{eqnarray}}
\newcommand{\ba}{\begin{Eqnarray}}
\newcommand{\ea}{\end{Eqnarray}}
\newcommand{\be}{\begin{equation}}
\newcommand{\ee}{\end{equation}}
\newcommand{\bal}{\begin{aligned}}
\newcommand{\eal}{\end{aligned}}
\newcommand{\bea}{\begin{eqnarray}}
\newcommand{\eea}{\end{eqnarray}}
\newcommand{\ben}{\begin{enumerate}}
\newcommand{\een}{\end{enumerate}}
\newcommand{\bit}{\begin{itemize}}
\newcommand{\eit}{\end{itemize}}
\newcommand{\bde}{\begin{widetext}}
\newcommand{\ede}{\end{widetext}}

\renewcommand{\[}{\left[}

\def\lsim{\mathrel{\rlap{\lower4pt\hbox{\hskip1pt$\sim$}}
    \raise1pt\hbox{$<$}}}
\def\gsim{\mathrel{\rlap{\lower4pt\hbox{\hskip1pt$\sim$}}
    \raise1pt\hbox{$>$}}}

\def\3211{$\mathrm{SU(3) \otimes SU(2)_L \otimes U(1)_R \otimes U(1)_{B-L}}$ }
\def\321{$\mathrm{SU(3) \otimes SU(2) \otimes U(1)}$ }
\def\422{$\mathrm{SU(4) \otimes SU(2) \otimes SU(2)_R}$ }

\newcommand{\mathsym}[1]{{}}

\title{

Dynamical scotogenic generation of the Linear and Inverse seesaws
}
\vspace*{1.8cm}

\author[a]{Asmaa Abada,}
\affiliation[a]{Pôle Théorie, Laboratoire de Physique des 2 Infinis Irène Joliot-Curie (UMR 9012), \\
  CNRS/IN2P3, 15 Rue Georges Clemenceau, 91400 Orsay, France}
\emailAdd{asmaa.abada@ijclab.in2p3.fr}

\author[b,c,d]{A.~E.~Cárcamo~Hernández}
\affiliation[b]{Universidad Técnica Federico Santa María, Casilla 110-V, Valparaíso, Chile}
\affiliation[c]{Centro Científico-Tecnológico de Valparaíso, Casilla 110-V, Valparaíso, Chile}
\affiliation[d]{Millennium Institute for Subatomic Physics at the High-Energy Frontier, SAPHIR, Chile}
\emailAdd{antonio.carcamo@usm.cl}

\author[a]{Salvador Urrea}
\emailAdd{salvador.urrea@ijclab.in2p3.fr}

\abstract{We propose an economical model in which the tiny active neutrino masses arise from an interplay of linear and inverse seesaw mechanisms. The Standard Model is extended by a local $U(1)'$ gauge symmetry and discrete $\mathbb{Z}_{3}\otimes\mathbb{Z}_{4}$ symmetries, together with gauge-singlet scalars and neutral leptons. Owing to the preserved discrete symmetries after spontaneous symmetry breaking, the linear and inverse seesaw mechanisms are dynamically generated at the two-loop level, while the same symmetries ensure the stability of both scalar and fermionic dark matter candidates. One of the distinctive features of the model is a fermionic dark matter candidate whose mass is generated at one loop, whereas scalar dark matter masses arise at tree level. The model satisfies current constraints from neutrino oscillation data, dark matter direct detection, invisible Higgs decays, $Z'$ searches, and charged lepton flavor violation, in addition we also discuss predictions for muonium states. The outcome of our analysis is that the inverse seesaw contribution dominates over the linear
one, suggesting  that atmospheric neutrino mass squared splitting arises from
the inverse seesaw mechanism, whereas the solar one is generated from
the linear seesaw. Finally, our model offers an explanation of the hierarchy between the atmospheric
and solar neutrino mass squared splittings, in addition to the smallness of active neutrino
masses, feature not presented in many low-scale seesaw models. In addition, our parameter-
space scan shows a slight preference for the normal neutrino mass ordering.}


\begin{document}

\maketitle

\section{Introduction}

The existence of dark matter and neutrino oscillation phenomena are strong indications of beyond Standard Model (BSM) physics, as the Standard Model (SM) provides neither a dark matter candidate nor a mechanism for generating active neutrino masses. 
The connection between the dark matter problem and the tiny active neutrino masses, together with the observed lepton mixing, offers an interesting and challenging avenue to explore.

While a minimal extension of the SM that introduces right-handed Majorana neutrinos ($\nu_R$) can generate active neutrino masses via the seesaw mechanism~\cite{Minkowski:1977sc, Yanagida:1979as, Glashow:1979nm, Mohapatra:1979ia, GellMann:1980vs, Schechter:1980gr, Schechter:1981cv}, accommodating sub-eV neutrino masses~\cite{Aker:2021gma} requires, in the canonical type-I seesaw realization, either a high lepton-number--violating (LNV) scale or tiny Yukawa couplings if implemented at low scale. In either case, the phenomenology of this mechanism is extremely suppressed. The resulting mixing between active and heavy neutrinos is very small, leading to rates for charged lepton flavor-violating (cLFV) processes that are far below current experimental sensitivity. An alternative that allows for large Yukawa couplings with a comparatively low seesaw scale is to introduce additional sterile fermions together with an approximate lepton number symmetry. 
This is the case in the inverse seesaw (ISS)~\cite{Wyler:1982dd, Mohapatra:1986bd, GonzalezGarcia:1988rw, Akhmedov:1995ip, Akhmedov:1995vm, Malinsky:2009df, Abada:2014vea} or the linear seesaw (LSS)~\cite{Barr:2003nn, Malinsky:2005bi} realizations.

Additionally, to generate tree-level Majorana masses for the active neutrinos, low-scale seesaw models introduce a new scale that explicitly breaks lepton number symmetry, either through a Majorana mass term (as in the ISS) or via an additional Yukawa interaction for the new fermion singlets (as in the LSS). In these scenarios, the LNV parameters are {\it ad hoc} and are assumed to be small, a hypothesis that, despite being technically natural and therefore stable under radiative corrections, lacks a more fundamental motivation.
Possible explanations for the origin of these small LNV parameters have been proposed in several models~\cite{Ma:2009gu, Bazzocchi:2010dt, Law:2012mj, Fraser:2014yha, Ahriche:2016acx, Das:2012ze, Das:2017ski, Das:2019pua, CarcamoHernandez:2013krw, CarcamoHernandez:2017owh, CarcamoHernandez:2018hst, CarcamoHernandez:2018iel, Bertuzzo:2018ftf, Mandal:2019oth, CarcamoHernandez:2019eme, CarcamoHernandez:2019pmy, CarcamoHernandez:2019vih, CarcamoHernandez:2019lhv, Hernandez:2021uxx, Hernandez:2021xet, Hernandez:2021kju, Nomura:2021adf, Hernandez:2021mxo, Abada:2021yot, Abada:2023zbb, Bonilla:2023egs, Bonilla:2023wok, Binh:2024lez, CarcamoHernandez:2024hll, Huong:2025uwx, Benitez-Irarrazabal:2025atb}.

 \medskip

 In this work we explore the possibility of explaining the origin of the tiny masses of the
active neutrinos (and the lepton mixing matrix) from the interplay of both linear and inverse seesaw mechanisms. The associated small LNV couplings of the LSS and ISS are dynamically generated at the two-loop level, which naturally justifies their smallness. In the model we propose, the SM gauge sector is extended by a local $U(1)'$ symmetry and by discrete $\mathbb{Z}_{3}\otimes \mathbb{Z}_{4}$ symmetries, while the particle content is enlarged with SM-gauge-singlet scalars and neutral leptons. A key feature of the model is the dynamical generation of the LNV parameters of the LSS and ISS mechanisms at two-loop level, arising from the spontaneous breaking of the local $U(1)'$ and the $\mathbb{Z}_{4}$ symmetries. This setup provides viable DM candidates, originating either from the sterile neutrino sector or from the extended scalar sector, which are in fact the particles responsible for generating both the inverse and linear seesaw mechanisms.
 
Many extensions of the SM that propose a radiative mass generation for the (new) neutral Majorana lepton masses rely on a preserved discrete symmetry that prevents tree-level mass contributions but allows them to appear at loop level~\cite{Balakrishna:1988ks, Ma:1988fp, Ma:1989ys, Ma:1990ce, Ma:1998dn, Tao:1996vb, Ma:2006km, Gu:2007ug, Ma:2008cu, Hirsch:2013ola, Aranda:2015xoa, Restrepo:2015ura, Longas:2015sxk, Fraser:2015zed, Fraser:2015mhb, Wang:2015saa, Arbelaez:2016mhg, vonderPahlen:2016cbw, Nomura:2016emz, Kownacki:2016hpm, Nomura:2017emk, Nomura:2017vzp, Bernal:2017xat, Wang:2017mcy, Bonilla:2018ynb, Calle:2018ovc, Avila:2019hhv, CarcamoHernandez:2018aon, Alvarado:2021fbw, Arbelaez:2022ejo, Cepedello:2022xgb, CarcamoHernandez:2022vjk, Leite:2023gzl}. These models naturally link dark matter and neutrino mass generation. In most cases, neutrino masses arise at one loop, which often necessitates either tiny Yukawa couplings, comparable to that of the electron, or very small mass splittings between the CP-even and CP-odd components of the scalar seesaw mediators.

On the other hand, models featuring linear~\cite{Wang:2015saa, CarcamoHernandez:2020pnh, CarcamoHernandez:2021tlv, CarcamoHernandez:2023atk, Batra:2023bqj, CarcamoHernandez:2024edi} or inverse~\cite{Mohapatra:1986bd, Malinsky:2005bi, Malinsky:2009df, Abada:2014vea, Ma:2009gu, Mandal:2019oth, Dev:2012sg, Guo:2012ne, Law:2012mj, Baldes:2013eva, Bazzocchi:2010dt, CarcamoHernandez:2019lhv, Hernandez:2021xet, Hernandez:2021kju, Abada:2021yot, Bonilla:2023egs, Bonilla:2023wok, Abada:2023zbb, Binh:2024lez, Gomez-Izquierdo:2024apr, Huong:2025uwx} seesaw mechanisms for the generation of light (active) neutrino masses offer an interesting and testable explanation for their smallness. These models provide sizeable active–heavy neutrino mixings, allowing cLFV processes to occur at rates within experimental reach, making linear and inverse seesaw constructions testable in cLFV experiments. Furthermore, when radiative mechanisms are implemented in linear and/or inverse seesaw models, the resulting scotogenic scenarios establish a connection between dark matter and neutrino mass generation, as some neutral seesaw messengers can serve as dark matter candidates. Their stability is ensured by discrete symmetries, which are also essential for the implementation of radiative seesaw mechanisms. Dark matter candidates in these models can annihilate into pairs of SM particles or other BSM states, successfully reproducing the observed dark matter relic abundance within suitable regions of parameter space. In addition, models embedding linear and inverse seesaw mechanisms can provide a viable framework for resonant leptogenesis due to the small mass splitting between the heavy pseudo-Dirac neutral leptons, since the physical heavy neutrinos form pseudo-Dirac pairs with a mass gap proportional to the small LNV parameters. All these arguments add up to reinforce the motivation to find a minimalist but testable scenario that can solve these problems with the same Lagrangian interactions.

In more detail, our model consists of an extension of the Inert Doublet Model, where the tiny masses of the active neutrinos are generated from an interplay of linear and inverse seesaw mechanisms—referred to as ``LISS''—both generated at the two-loop level. The SM gauge symmetry is augmented by the inclusion of a local $U(1)^{\prime}$ and the discrete symmetries $\mathbb{Z}_{3}\otimes \mathbb{Z}_{4}$, while the SM particle content is enlarged by several gauge-singlet scalars and neutral leptons. The local $U(1)^{\prime}$ and the $\mathbb{Z}_{4}$ symmetries are spontaneously broken, whereas the $\mathbb{Z}_{3}$ symmetry remains exact. We assume that the $U(1)^{\prime}$ is spontaneously broken down to a residual $\mathbb{Z}_{2}$ discrete symmetry, the preserved $\mathbb{Z}_{2}\times\mathbb{Z}_{3}$ symmetries being essential to ensure the radiative nature of the linear and inverse seesaw mechanisms at the two-loop level. Under these assumptions, our model is anomaly-free, both gauge and gravitational. In addition to dynamically generating the LISS and thus small masses for the active neutrinos, our model provides a scotogenic origin for the LNV parameters of the ISS and LSS. The model therefore supplies two dark matter candidates, one scalar and one fermionic, both stable thanks to the residual $\mathbb{Z}_{2}\times \mathbb{Z}_{3}$ symmetry.  A notable feature is the difference in their mass generation. The fermionic candidate, identified as the lightest neutral lepton carrying a non-trivial $\mathbb{Z}_{3}$ charge, acquires mass radiatively at the one-loop level. All scalar dark matter masses, in contrast, arise at tree level. It yields sizeable rates for cLFV observables and among these, we evaluate the contributions to the decay rates of $\ell_\alpha\to \ell_\beta\ell_\beta\ell_\rho$, and to the radiative processes $\mu \to e\gamma$, $\tau \to \mu \gamma$, and $\tau \to e\gamma$, as well as muon–electron conversion in nuclei, $\mathrm{CR}(\mu-e, \mathrm{N})$. We furthermore study the impact of our model on the muonium (Mu) state, including cLFV decays and muonium–antimuonium oscillations, providing predictions and a phenomenological analysis. We already anticipate highlighting that in this scenario, the constraint from neutrino data on lepton mixing implies that the dominant contribution to the cLFV amplitudes (boxes and penguins) is essentially due to the scalar sector. Furthermore, in the fermionic DM scenario, constraints arising from the LUX experiment~\cite{LZ:2024zvo} set the ratio $\frac{M_{Z^{\prime}}}{g_{X}}$ to be larger than about $40$ TeV for a $10$ GeV fermionic dark matter mass.

The paper is organized as follows: in Section~\ref{sec:scotogenic_model} we provide a detailed description of the model with its symmetries and field content. We also discuss the stability condition of the scalar potential and its consequences. The dynamical generation of the complete mass matrix of the neutral leptons at the two-loop level is derived in Section~\ref{Sec:mass-generation}; the generation of active light neutrino masses and the lepton mixing is also detailed in this section. Section~\ref{sec:pheno} is devoted to the phenomenological study of our model with high-intensity observables, mainly cLFV ones, collider searches, as well as current constraints. For this, we present a detailed calculation of the observables we consider. Our results are presented and discussed in Section~\ref{sec:results}. We finally summarize our findings in Section~\ref{Sec:conclusions}. The Appendices provide supplementary details on anomaly cancellation, the stability and unitarity conditions of the model, and the explicit form factors for cLFV observables.

\section{The scotogenic LISS model}\label{sec:scotogenic_model}

In this section, we present the two-loop level scotogenic inverse and linear seesaw model we propose.

\subsection{Symmetries and particles}\label{Sec:model}

The model corresponds to a minimal extension of the inert
doublet model, where the scalar field content is augmented by the inclusion
of the gauge singlet fields $\varphi $, $\xi $, $\chi $, $\rho $, $\sigma $
and 
the fermion sector is enlarged with 
 Majorana neutrinos $\nu _{R}$, $N_{R}$,  $\Psi _{L}$, $\Psi_{R}$ and $\tilde{\Psi}_{R}$, and 
  heavy Dirac neutral leptons $\Omega _{R}$, $\Omega _{L}$. 
Regarding the gauge group, we extend the SM 
with a local  $U(1)^{\prime }$  and  discrete $\mathbb{Z}%
_{3}\otimes \mathbb{Z}_{4}$  symmetries. The  scalar potential of our model 
yields a tree-level instability forming vacuum expectation values (VEVs) of 
$\langle \phi \rangle =v_{\phi }$, being  $\phi$ the SM Higgs doublet, $\langle \sigma \rangle =v_{\sigma },\langle
\rho \rangle =v_{\rho }$ that trigger the following spontaneous symmetry
breaking (SSB) pattern: 
\begin{align}
& SU(3)_{C}\otimes SU(2)_{L}\otimes U(1)_{Y}\otimes U(1)^{\prime }\otimes 
\mathbb{Z}_{3}\otimes \mathbb{Z}_{4}  \\
& \hspace{20mm}\Downarrow v_{\sigma },v_{\rho }  \notag \\
& SU(3)_{C}\otimes SU(2)_{L}\otimes U(1)_{Y}\otimes 
\underbrace{\widetilde{\mathbb{Z}}_{6}}_{\text{remnant of }U(1)'}
\otimes \mathbb{Z}_{3}\otimes 
\underbrace{\mathbb{Z}_{2}}_{\mathbb{Z}_{4}\to\mathbb{Z}_{2}} \\
& \hspace{20mm}\Downarrow v_{\phi }  \notag \\
& SU(3)_{C}\otimes U(1)_{\text{em}}\ \otimes 
\widetilde{\mathbb{Z}}_{6}\otimes \mathbb{Z}_{3}\otimes \mathbb{Z}_{2}\,,
\end{align}

After spontaneous symmetry breaking, the local $%
U(1)^{\prime }$ symmetry is spontaneously broken down to a conserved $\widetilde{\mathbb{Z}}_6$ discrete group which has as subgroup the preserved matter
parity $\widetilde{\mathbb{Z}}_{2}\subset \widetilde{\mathbb{Z}}_6$ symmetry with the $\widetilde{\mathbb{Z}}%
_{2}$ charges of the fields defined as $\left( -1\right) ^{3Q_{U(1)^{\prime
}}+2s}$, where $Q_{U(1)^{\prime
}}$ is the charge under ${U(1)^{\prime }}$ and $s$ is the spin of the particle%
\footnote{Strictly speaking, the charge assignment for the preserved subgroup $\widetilde{\mathbb{Z}}_{2}\subset\widetilde{\mathbb{Z}}_{6}$ is $(-1)^{3Q_{U(1)^{\prime}}}$. However, Lorentz invariance ensures that only an even number of fermionic fields can appear in operators, making both assignments physically equivalent.}.  
Finally, the $\mathbb{Z}_{4}$ is broken to another independent preserved $\mathbb{Z}_{2}$ symmetry. The charge assignments under the extended $SU(3)_{C}\otimes
SU(2)_{L}\otimes U(1)_{Y}\otimes U(1)^{\prime }\otimes \mathbb{Z}_{3}\otimes 
\mathbb{Z}_{4}$ gauge group and the residual $\widetilde{\mathbb{Z}}_{2}$ symmetry for all particles of the model are
provided in Table~\ref{model}.

\begin{table}[h!]
\renewcommand{\arraystretch}{1.3} 
\centering
\resizebox{\textwidth}{!}{%
\begin{tabular}{|c|c|c|c|c|c|c|c|c|c|c|c|c|c|c|c|c|c|c|c|}
\hline
Field & $q_{iL}$ & $u_{iR}$ & $d_{iR}$ & $l_{iL}$ & $l_{iR}$ & $\nu _{R}$ & $N_{R}$ & $\Psi _{L}$ & $\Psi _{R}$ & $\tilde{\Psi}_{R}$ & $\Omega _{L}$ & $\Omega _{R}$ & $\phi $ & $\eta $ & $\varphi $ & $\xi $ & $\chi $ & $\sigma $ & $\rho $ \\ \hline\hline
$SU(3)_{C}$ & $\mathbf{3}$ & $\mathbf{3}$ & $\mathbf{3}$ & $\mathbf{1}$ & $\mathbf{1}$ & $\mathbf{1}$ & $\mathbf{1}$ & $\mathbf{1}$ & $\mathbf{1}$ & $\mathbf{1}$ & $\mathbf{1}$ & $\mathbf{1}$ & $\mathbf{1}$ & $\mathbf{1}$ & $\mathbf{1}$ & $\mathbf{1}$ & $\mathbf{1}$ & $\mathbf{1}$ & $\mathbf{1}$ \\ \hline
$SU(2)_{L}$ & $\mathbf{2}$ & $\mathbf{1}$ & $\mathbf{1}$ & $\mathbf{2}$ & $\mathbf{1}$ & $\mathbf{1}$ & $\mathbf{1}$ & $\mathbf{1}$ & $\mathbf{1}$ & $\mathbf{1}$ & $\mathbf{1}$ & $\mathbf{1}$ & $\mathbf{2}$ & $\mathbf{2}$ & $\mathbf{1}$ & $\mathbf{1}$ & $\mathbf{1}$ & $\mathbf{1}$ & $\mathbf{1}$ \\ \hline
$U(1)_{Y}$ & $\tfrac{1}{6}$ & $\tfrac{2}{3}$ & $-\tfrac{1}{3}$ & $-\tfrac{1}{2}$ & $-1$ & $0$ & $0$ & $0$ & $0$ & $0$ & $0$ & $0$ & $\tfrac{1}{2}$ & $-\tfrac{1}{2}$ & $0$ & $0$ & $0$ & $0$ & $0$ \\ \hline
$U(1)^{\prime }$ & $\tfrac{1}{3}$ & $\tfrac{1}{3}$ & $\tfrac{1}{3}$ & $-1$ & $-1$ & $-1$ & $1$ & $1$ & $-1$ & $-1$ & $-2$ & {$-2$} & $0$ & $0$ & $0$ & $-1$ & {$3$} & $2$ & $0$ \\ \hline
$\mathbb{Z}_{3}$ & $0$ & $0$ & $0$ & $0$ & $0$ & $0$ & $0$ & $1$ & $1$ & {$0$} & $-1$ & $-1$ & $0$ & $-1$ & $1$ & $1$ & $-1$ & $0$ & $0$ \\ \hline
$\mathbb{Z}_{4}$ & $0$ & $0$ & $0$ & $0$ & $0$ & $0$ & $0$ & $0$ & $0$ & {$1$} & $1$ & {$1$} & $0$ & $0$ & $0$ & {$1$} & {$-1$} & {$2$} & {$2$} \\ \hline
$\widetilde{\mathbb{Z}}_{6}$ & $1$ & $1$ & $1$ & $3$ & $3$ & $3$ & $3$ & $3$ & $3$ & $3$ & $0$ & $0$ & $0$ & $0$ & $0$ & $3$ & $3$ & $0$ & $0$ \\ \hline
$\widetilde{\mathbb{Z}}_{2}$ & $0$ & $0$ & $0$ & $0$ & $0$ & $0$ & $0$ & $0$ & $0$ & $0$ & $1$ & $1$ & $0$ & $0$ & $0$ & $1$ & $1$ & $0$ & $0$ \\ \hline
$\mathbb{Z}_{2}$ & $0$ & $0$ & $0$ & $0$ & $0$ & $0$ & $0$ & $0$ & $0$ & {$1$} & $1$ & $1$ & $0$ & $0$ & $0$ & $1$ & $1$ & $0$ & $0$ \\ \hline
\end{tabular}%
}
\caption{Particle charge assignments under 
$SU(3)_{C}\times SU(2)_{L}\times U(1)_{Y}\times U(1)^{\prime }\times \mathbb{Z}_{3}\times \mathbb{Z}_{4}$ 
and the residual discrete symmetries after spontaneous breaking. The model itself is summarized on the 6 first rows of the table. Here $i=1,2,3$ and for completeness $\widetilde{\mathbb{Z}}_{6}$ is the remnant of $U(1)'$, $\widetilde{\mathbb{Z}}_{2}$ is its matter-parity subgroup, 
and $\mathbb{Z}_{2}$ is the subgroup preserved after $\mathbb{Z}_{4}$ breaking.}
\label{model}
\end{table}

The spontaneous breaking of the $U(1)^{\prime}$ and $\mathbb{Z}_{2}$ symmetries occurs at the TeV scale, triggered by the vacuum expectation values (VEVs) of the scalar fields $\sigma$ and $\rho$, respectively. The other scalar fields $\eta$, $\xi$, $\chi$, and $\varphi$, which carry non–trivial charges under $\widetilde{\mathbb{Z}}_{2}\otimes\mathbb{Z}_{3}$, do not acquire VEVs. This ensures that the $\widetilde{\mathbb{Z}}_{2}\otimes\mathbb{Z}_{3}$ symmetry remains exact. Preserving this symmetry is essential both to guarantee the radiative origin of the linear and inverse seesaw mechanisms and to ensure the stability of the scalar and fermionic dark matter candidates.

Regarding the dynamical generation of lepton number violation, the one–loop radiative linear and inverse seesaw mechanisms are forbidden by the symmetries of the model implying 
the absence of tree–level Dirac and Majorana mass terms. In particular, the Dirac mass term $M_{\Psi}^{D}\,\overline{\Psi}_{L}\Psi_{R}$ and the Majorana mass term $M_{\Psi}^{M}\,\Psi_{R}\,\overline{\Psi_{R}^{C}}$ are both absent: the Majorana term is forbidden by the local $U(1)^{\prime}$ symmetry, whereas the Dirac term is excluded by the $\mathbb{Z}_{3}$ symmetry. In other words, these terms cannot be generated dynamically at tree level because the Yukawa couplings $\overline{\Psi}_{L}\sigma\,\Psi_{R}$ and $\Psi_{R}\sigma\,\overline{\Psi_{R}^{C}}$ are not invariant under $U(1)^{\prime}$ and are simultaneously forbidden by the $\mathbb{Z}_{3}$ and $\mathbb{Z}_{4}$ symmetries. Consequently, the preserved $\mathbb{Z}_{3}$ and $\mathbb{Z}_{4}$ symmetries play a crucial role in preventing one–loop active neutrino masses from arising through the linear and inverse seesaw mechanisms.

Moreover, both $U(1)^{\prime}$ and $\mathbb{Z}_{4}$ forbid the Yukawa interaction $N_{R}\sigma\,\overline{N_{R}^{C}}$, which would otherwise induce a tree–level inverse seesaw contribution. It is important to note that the Majorana mass term 

$ M_{\Psi}\,\Psi_{R}\,\overline{\Psi_{R}^{C}}$ only appears at the one–loop level, generated via the virtual exchange of the Majorana leptons $\Omega_{R}$ together with the real and imaginary components of the singlet scalars $\xi$ and $\chi$. In contrast, the Majorana mass for $\tilde{\Psi}_{R}$ is already present at tree level, arising from its coupling to $\sigma$. Nevertheless, due to the nontrivial $\mathbb{Z}_{4}$ charge of $\tilde{\Psi}_{R}$, this contribution does not participate neither in the linear nor in the inverse seesaw mechanisms. The field $\tilde{\Psi}_{R}$ is required instead to ensure anomaly cancellation of the $U(1)^{\prime}$ gauge symmetry.

Finally, as will be shown later, the scalar fields $\sigma$ and $\rho$ are necessary to close the scalar line of the internal loop in Fig.~\ref{epdiagram} through the quartic interaction $\lambda_{\xi \sigma \chi \rho}\Bigl(\xi \sigma^* \chi \rho+\text {h.c.}\Bigr)$, while the external loop is completed by the trilinear interaction $\Bigl( A_{\phi\eta\varphi}\,\phi\,\eta\,\varphi + \text{h.c.} \Bigr)$.

Beyond its particle-physics implications, the spontaneous breaking of the discrete $Z_4$ symmetry  leads to the formation of cosmological domain walls (DWs)~\cite{Vilenkin:1981zs,Kibble:1982dd,Vilenkin:1984ib}. If persistent, these structures could eventually dominate the energy budget of the universe, disrupting primordial nucleosynthesis (BBN). A common resolution to this problem, assuming DWs form after inflation, is to introduce a small soft breaking parameter (e.g. $A_{\rho}\rho^2$) that renders them unstable (see e.g. Ref.~\cite{Lazarides:2018aev} for other approaches). In this context, the soft-breaking mass terms provide a bias potential $\Delta V_{\rm bias}$ that lifts the degeneracy between distinct vacua. This creates a pressure differential across the DWs, prompting their collapse once the breaking is sufficiently large~\cite{Matsuda:1998ms,Dvali:1995cc,Abel:1995wk,Rai:1992xw,Nakayama:2016gxi,Saikawa:2017hiv,Lazarides:2018aev}. The resulting decay rate is proportional to the magnitude of the soft-breaking parameters, while the DW lifetime $\tau_{\rm DW}$ is inversely proportional. To ensure decay occurs before BBN and before DWs dominate the universe ($\tau_{\rm DW} \lesssim t_{\rm DW}\sim M_P^2/\sigma_\omega$), the bias must be large enough to satisfy $\tau_{\rm DW}\approx\sigma_\omega/\Delta V_{\rm bias}\lesssim 0.1$~s, where the wall tension is $\sigma_\omega\sim v_\rho^3$.

For a bias potential of the form $\Delta V_{\rm bias}=\mu_{\rm bias}^3v_\rho$, these constraints translate into a lower bound on the symmetry-breaking scale \cite{CarcamoHernandez:2023oeq}:
\begin{align}
\frac{A_\rho}{{\rm TeV}}\gtrsim {\rm max }\left[10^{-5}\left(\frac{v_\rho}{{\rm TeV}}\right)^{2/3}, 10^{-5}\left(\frac{v_\rho}{{\rm TeV}}\right)^{5/3}\right].
\end{align}
This lower bound is sufficiently small, implying that including soft-breaking parameters would lead to subleading contributions to neutrino mass generation compared to the two-loop inverse and linear seesaw mechanisms considered here; therefore, they are neglected in this work.

The collapse of the DW network can generate a stochastic gravitational wave background, potentially observable by current and future experiments~\cite{Hiramatsu:2013qaa,Saikawa:2017hiv}. A precise determination of the wall tension and bias potential requires a sophisticated analysis of the model's finite-temperature effective potential, as undertaken in works like \cite{Sassi:2023cqp,Battye:2020sxy,Chen:2020soj}. Such a detailed study is reserved for future publication.

\subsection{The scalar potential}\label{Sec:scalar-potential}
The most general scalar potential invariant under the symmetries of the model, is given by:
\begin{align}\label{eq:pot}
V &= 
-\mu_{\phi}^{2}(\phi^{\dagger}\phi)
-\mu_{\sigma}^{2}(\sigma^{\ast}\sigma)
-\mu_{\rho}^{2}\rho^{2}
+\mu_{\eta}^{2}(\eta^{\dagger}\eta)
+\mu_{\varphi}^{2}(\varphi^{\ast}\varphi)
+\mu_{\xi}^{2}(\xi^{\ast}\xi)
+\mu_{\chi}^{2}(\chi^{\ast}\chi) \nonumber\\[1mm]
&\quad
+ \bigl(A_{\phi\eta\varphi}\,\phi\,\eta\,\varphi + \text{h.c.}\bigr)
+ \lambda_{\phi}(\phi^{\dagger}\phi)^2
+ \lambda_{\sigma}(\sigma^{\ast}\sigma)^2
+ \lambda_{\rho}\rho^4
+ \lambda_{\eta}(\eta^{\dagger}\eta)^2
+ \lambda_{\varphi}(\varphi^{\ast}\varphi)^2 \nonumber\\[1mm]
&\quad
+ \lambda_{\xi}(\xi^{\ast}\xi)^2
+ \lambda_{\chi}(\chi^{\ast}\chi)^2
+ \lambda_{\phi\sigma}(\phi^{\dagger}\phi)(\sigma^{\ast}\sigma)
+ \lambda_{\phi\rho}(\phi^{\dagger}\phi)\rho^2
+ \lambda_{\rho\sigma}\rho^2(\sigma^{\ast}\sigma)
+ \lambda_{\eta\varphi}(\eta^{\dagger}\eta)(\varphi^{\ast}\varphi) \nonumber\\[1mm]
&\quad
+ \lambda_{\eta\xi}(\eta^{\dagger}\eta)(\xi^{\ast}\xi)
+ \lambda_{\eta\chi}(\eta^{\dagger}\eta)(\chi^{\ast}\chi)
+ \lambda_{\varphi\xi}(\varphi^{\ast}\varphi)(\xi^{\ast}\xi)
+ \lambda_{\varphi\chi}(\varphi^{\ast}\varphi)(\chi^{\ast}\chi)
+ \lambda_{\xi\chi}(\xi^{\ast}\xi)(\chi^{\ast}\chi) \nonumber\\[1mm]
&\quad
+ \lambda_{\phi\eta}(\phi^{\dagger}\phi)(\eta^{\dagger}\eta)
+ \lambda'_{\phi\eta}(\phi\,\eta)(\eta^{\dagger}\phi^{\dagger})
+ \lambda_{\phi\varphi}(\phi^{\dagger}\phi)(\varphi^{\ast}\varphi)
+ \lambda_{\phi\xi}(\phi^{\dagger}\phi)(\xi^{\ast}\xi)
+ \lambda_{\phi\chi}(\phi^{\dagger}\phi)(\chi^{\ast}\chi) \nonumber\\[1mm]
&\quad
+ \lambda_{\sigma\eta}(\sigma^{\ast}\sigma)(\eta^{\dagger}\eta)
+ \lambda_{\sigma\varphi}(\sigma^{\ast}\sigma)(\varphi^{\ast}\varphi)
+ \lambda_{\sigma\xi}(\sigma^{\ast}\sigma)(\xi^{\ast}\xi)
+ \lambda_{\sigma\chi}(\sigma^{\ast}\sigma)(\chi^{\ast}\chi) \nonumber\\[1mm]
&\quad
+ \lambda_{\rho\eta}\rho^2(\eta^{\dagger}\eta)
+ \lambda_{\rho\varphi}\rho^2(\varphi^{\ast}\varphi)
+ \lambda_{\rho\xi}\rho^2(\xi^{\ast}\xi)
+ \lambda_{\rho\chi}\rho^2(\chi^{\ast}\chi)
+ \bigl(\lambda_{\xi\sigma\chi\rho}\,\xi\sigma^{\ast}\chi\rho + \text{h.c.}\bigr).
\end{align}

\noindent Observe 
that the $\rho$ field can be assumed real without loss of generality. 

We are interested in a scenario where the fields $\phi$, $\sigma$, and $\rho$ develop VEVs ($v_\phi$, $v_\sigma$, and $v_\rho$, respectively), while the fields $\eta$, $\varphi$, $\xi$, and $\chi$ do not acquire any VEV, since the latter have non trivial charges under the preserved $\widetilde{\mathbb{Z}}_{2}\otimes\mathbb{Z}_{3}$ symmetry. 
Thus, in the scalar potential, we take $-\mu_\phi^2$, $-\mu_\sigma^2$, and $-\mu_\rho^2$ as negative, whereas $\mu_{\eta}^{2}$, $\mu_{\varphi}^{2}$, $\mu_{\xi}^{2}$ and $\mu_{\chi}^{2}$ are taken to be positive. We adopt a CP-conserving scenario, which forbids mixing between CP-even and CP-odd scalars. This, in turn, implies that the trilinear coupling $A_{\phi\eta\varphi}$ and all quartic terms are real. 
Further restrictions on the parameters arise from potential stability and unitarity, detailed in Appendix~\ref{sec:stability-unitarity}. The scalar fields are defined as follows:

\begin{equation}
\begin{array}{c@{\hspace{2cm}}c}
\begin{aligned}
\phi &= \begin{pmatrix}\phi^{+}\\[2mm]\frac{1}{\sqrt{2}}\Bigl(v_{\phi}+h+i\,\phi_Z\Bigr)\end{pmatrix}, \\[2mm]
\sigma &= \frac{1}{\sqrt{2}}\Bigl(v_\sigma+\tilde{\sigma}+i\,\sigma_{Z'}\Bigr),\\[2mm]
\chi &= \frac{1}{\sqrt{2}}\chi_R+i\,\frac{1}{\sqrt{2}}\chi_I, \\[2mm]
\rho &= v_\rho+\tilde{\rho}, \\[2mm]
\end{aligned}
&
\begin{aligned}
\eta &= \begin{pmatrix}\frac{1}{\sqrt{2}}\Bigl(\eta^0_R+i\,\eta^0_I\Bigr)\\[2mm]\eta^{-}\end{pmatrix}, \\[2mm]
\varphi &= \frac{1}{\sqrt{2}}\varphi_R+i\,\frac{1}{\sqrt{2}}\varphi_I, \\[2mm]
\xi &= \frac{1}{\sqrt{2}}\xi_R+i\,\frac{1}{\sqrt{2}}\xi_I.
\end{aligned}
\end{array}
\end{equation}
where $v_\phi$, $v_\sigma$, and $v_\rho$ denote the VEVs responsible for breaking the electroweak symmetry, $U(1)'$, and $\mathbb{Z}_4$, respectively. The scalar remnants of spontaneous symmetry breaking are $h$, $\tilde{\sigma}$, $\tilde{\rho}$, $\eta^0$, $\eta^{-}$, $\varphi$, $\xi$, and $\chi$. Moreover, $\phi^{\pm}$, $\phi_Z$, and $\sigma_{Z'}$ are the Goldstone bosons corresponding to the longitudinal components of the gauge fields $W^\pm$, $Z$, and $Z'$, respectively. Minimizing the scalar potential by setting its derivatives with respect to the VEVs to zero yields the following conditions on the couplings defined in Eq.~(\ref{eq:pot}): 
\begin{align}
2\lambda_{\phi} v_\phi^2 + \lambda_{\phi\sigma} v_\sigma^2 + 2\lambda_{\phi\rho} v_\rho^2 + 2\mu_{\phi}^2 &=0 , \\
\lambda_{\phi\sigma} v_\phi^2 + 2\lambda_{\rho\sigma} v_\rho^2 + 2\lambda_{\sigma} v_\sigma^2 + 2\mu_\sigma^2 &= 0, \\
\lambda_{\phi\rho} v_\phi^2 + \lambda_{\rho\sigma} v_\sigma^2 + 4\lambda_{\rho} v_\rho^2 + 2\mu_\rho^2 &= 0.
\end{align}

After SSB there would be mixing among the scalars determined by the charges under $Z_3$ which remains unbroken and the remnant $\widetilde{\mathbb{Z}}_2$, dividing the scalars into three sectors, 
as follows: 
$(h,\tilde{\sigma},\tilde{\rho})$, $(\eta^0,\chi)$, and $(\xi,\varphi)$. Starting with the $\widetilde{\mathbb{Z}}_2$-even and $\mathbb{Z}_3$ neutral we have:  

\begin{equation}
-\mathcal{L}_\text{mass}^\text{$\widetilde{\mathbb{Z}}_2$-even,$\mathbb{Z}_3$-neutral } = \frac{1}{2} 
\begin{pmatrix} 
h & \tilde{\sigma} & \tilde{\rho} 
\end{pmatrix}
\begin{pmatrix}
2 \lambda_{\phi} v_\phi^2 & \lambda_{\phi\sigma} v_\phi v_\sigma & 2 \lambda_{\phi\rho} v_\phi v_\rho \\
\lambda_{\phi\sigma} v_\phi v_\sigma & 2 \lambda_{\sigma} v_\sigma^2 & 2 \lambda_{\rho\sigma} v_\sigma v_\rho \\
2 \lambda_{\phi\rho} v_\phi v_\rho & 2 \lambda_{\rho\sigma} v_\sigma v_\rho & 8 \lambda_{\rho} v_\rho^2
\end{pmatrix}
\begin{pmatrix} 
h \\ \tilde{\sigma} \\ \tilde{\rho} 
\end{pmatrix}.
\end{equation}

The quartic couplings $\lambda_{\phi\sigma}$ and $\lambda_{\phi\rho}$ govern the extent of mixing between $h$ and the additional scalars $\tilde{\sigma}$ and $\tilde{\rho}$, influencing deviations from the SM Higgs boson properties, which are tightly constrained by LHC Higgs measurements~\cite{ATLAS:2022vkf,CMS:2022dwd,ParticleDataGroup:2024cfk}. 
To align with these observations, we consider the alignment limit where $\lambda_{\phi\sigma} = \lambda_{\phi\rho} = 0$, ensuring $h$ closely mimics the SM Higgs. In this scenario, the masses of the eigenstates $\tilde{\sigma}^\prime$ and $\tilde{\rho}^\prime$ can be analytically determined, with the mixing defined as:

\begin{equation}
\left( \begin{array}{c} \tilde{\sigma}^\prime \\ \tilde{\rho}^\prime \end{array} \right) = 
\left( \begin{array}{cc} \cos\theta & -\sin\theta \\ \sin\theta & \cos\theta \end{array} \right)
\left( \begin{array}{c} \tilde{\sigma} \\ \tilde{\rho} \end{array} \right),
\end{equation}
where the mass eigenvalues and mixing angle are given by:
\begin{align}
m_{\tilde{\sigma}^\prime}^2 &= 2 \lambda_{\sigma} v_\sigma^2 + 8 \lambda_{\rho} v_\rho^2 + \sqrt{\left(2 \lambda_{\sigma} v_\sigma^2 - 8 \lambda_{\rho} v_\rho^2\right)^2 + 4 \left(2 \lambda_{\rho\sigma} v_\rho v_\sigma\right)^2}, \\
m_{\tilde{\rho}^\prime}^2 &= 2 \lambda_{\sigma} v_\sigma^2 + 8 \lambda_{\rho} v_\rho^2 - \sqrt{\left(2 \lambda_{\sigma} v_\sigma^2 - 8 \lambda_{\rho} v_\rho^2\right)^2 + 4 \left(2 \lambda_{\rho\sigma} v_\rho v_\sigma\right)^2}, \\
\tan 2\theta &= \frac{4 \lambda_{\rho\sigma} v_\rho v_\sigma}{2 \lambda_{\sigma} v_\sigma^2 - 8 \lambda_{\rho} v_\rho^2}.
\end{align}

For the $\widetilde{\mathbb{Z}}_2$-even scalars charged under $\mathbb{Z}_3$, the potential induces mixing among the scalar fields for both the imaginary and the real parts. The mass matrices for both real and imaginary components are presented compactly, differing only by the sign of the trilinear terms:
\begin{multline}
-\mathcal{L}_\text{mass}^{\widetilde{\mathbb{Z}}_2\text{-even},\,\mathbb{Z}_3\text{-charged}} = \\
\frac{1}{2} (\eta_{R/I},\, \varphi_{R/I})
\begin{pmatrix}
\mu_\eta^2 + \frac{1}{2} \lambda_{\phi\eta} v_\phi^2 + \frac{1}{2} \lambda_{\sigma\eta} v_\sigma^2 + \lambda_{\rho\eta} v_\rho^2 & \pm \frac{1}{\sqrt{2}} A_{\phi\eta\varphi} v_\phi \\
\pm \frac{1}{\sqrt{2}} A_{\phi\eta\varphi} v_\phi & \mu_\varphi^2 + \frac{1}{2} \lambda_{\phi\varphi} v_\phi^2 + \frac{1}{2} \lambda_{\sigma\varphi} v_\sigma^2 + \lambda_{\rho\varphi} v_\rho^2
\end{pmatrix}
\begin{pmatrix} \eta_{R/I} \\ \varphi_{R/I} \end{pmatrix},
\end{multline}
where $+$ applies to the real parts $(\eta_R, \varphi_R)$ and $-$ to imaginary ones $(\eta_I, \varphi_I)$. Diagonalization proceeds via the following rotation:
\begin{equation}
\begin{pmatrix} \eta_{R/I} \\ \varphi_{R/I} \end{pmatrix}
= R_{S_{R/I}} \begin{pmatrix} S_{1\,R/I} \\ S_{2\,R/I} \end{pmatrix} \text{ with }  R_{S_{R/I}} = 
\left( \begin{array}{cc} \cos \theta_{S_{R/I}} & \sin \theta_{S_{R/I}} \\ -\sin \theta_{S_{R/I}} & \cos \theta_{S_{R/I}} \end{array} \right), 
\label{eq:R}
\end{equation}
with the mixing angle given by:
\begin{equation}
\tan 2 \theta_{S_{R/I}} = \frac{ \pm \sqrt{2}\, A_{\phi\eta\varphi}\, v_\phi }{ \left( \mu_\eta^2 + \frac{1}{2} \lambda_{\phi\eta} v_\phi^2 + \frac{1}{2} \lambda_{\sigma\eta} v_\sigma^2 + \lambda_{\rho\eta} v_\rho^2 \right) - \left( \mu_\varphi^2 + \frac{1}{2} \lambda_{\phi\varphi} v_\phi^2 + \frac{1}{2} \lambda_{\sigma\varphi} v_\sigma^2 + \lambda_{\rho\varphi} v_\rho^2 \right) },
\end{equation}
and the eigenvalues as below:
\begin{align}
m_{S_{1{R/I}}}^2 &= \frac{1}{2} \left[ m_{\eta}^2 + m_{\varphi}^2 + \sqrt{ \left(m_{\eta}^2 - m_{\varphi}^2\right)^2 + 2 A_{\phi\eta\varphi}^2 v_\phi^2 } \right], \\
m_{S_{2{R/I}}}^2 &= \frac{1}{2} \left[ m_{\eta}^2 + m_{\varphi}^2 - \sqrt{ \left(m_{\eta}^2 - m_{\varphi}^2\right)^2 + 2 A_{\phi\eta\varphi}^2 v_\phi^2 } \right],
\end{align}
where $m_{\eta}^2 \equiv \mu_\eta^2 + \frac{1}{2} \lambda_{\phi\eta} v_\phi^2 + \frac{1}{2} \lambda_{\sigma\eta} v_\sigma^2 + \lambda_{\rho\eta} v_\rho^2,\quad
m_{\varphi}^2 \equiv \mu_\varphi^2 + \frac{1}{2} \lambda_{\phi\varphi} v_\phi^2 + \frac{1}{2} \lambda_{\sigma\varphi} v_\sigma^2 + \lambda_{\rho\varphi} v_\rho^2.$
Note that trilinear terms enter only off-diagonally and with opposite sign for real and imaginary components, this makes the masses degenerate and the mixing angle having opposite sign ($\theta_{S_{R}}=-\theta_{S_{I}}$).

Similarly for the $\widetilde{\mathbb{Z}}_2$-odd scalars charged under $\mathbb{Z}_3$, the mixing reads: 
\begin{multline}
\mathcal{L}_\text{mass}^{\widetilde{\mathbb{Z}}_2\text{-odd},\,\mathbb{Z}_3\text{-charged}} = \\
\frac{1}{2} (\xi_{R/I},\, \chi_{R/I})
\begin{pmatrix}
\mu_\xi^2 + \frac{1}{2} \lambda_{\phi\xi} v_\phi^2 + \frac{1}{2} \lambda_{\sigma\xi} v_\sigma^2 + \lambda_{\rho\xi} v_\rho^2 & \pm \lambda_{\xi \sigma \chi \rho} v_\rho v_\sigma \\
\pm \lambda_{\xi \sigma \chi \rho} v_\rho v_\sigma & \mu_\chi^2 + \frac{1}{2} \lambda_{\phi\chi} v_\phi^2 + \frac{1}{2} \lambda_{\sigma\chi} v_\sigma^2 + \lambda_{\rho\chi} v_\rho^2
\end{pmatrix}
\begin{pmatrix} \xi_{R/I} \\ \chi_{R/I} \end{pmatrix},
\end{multline}
diagonalized by:
\begin{equation}
\begin{pmatrix} \xi_{R/I} \\ \chi_{R/I} \end{pmatrix}
= \widetilde{R}_{S_{R/I}} \begin{pmatrix} S_{3\,R/I} \\ S_{4\,R/I} \end{pmatrix} 
\quad \text{with} \quad
\widetilde{R}_{S_{R/I}} =
\begin{pmatrix}
\cos \gamma_{S_{R/I}} & \sin \gamma_{S_{R/I}} \\
-\sin \gamma_{S_{R/I}} & \cos \gamma_{S_{R/I}}
\end{pmatrix}.
\label{eq:R_tilde}
\end{equation}
with mixing:
\begin{equation}
\tan 2 \gamma_{S_{R/I}} = \frac{ \pm 2 \lambda_{\xi \sigma \chi \rho} v_\rho v_\sigma }{ \left( \mu_\xi^2 + \frac{1}{2} \lambda_{\phi\xi} v_\phi^2 + \frac{1}{2} \lambda_{\sigma\xi} v_\sigma^2 + \lambda_{\rho\xi} v_\rho^2 \right) - \left( \mu_\chi^2 + \frac{1}{2} \lambda_{\phi\chi} v_\phi^2 + \frac{1}{2} \lambda_{\sigma\chi} v_\sigma^2 + \lambda_{\rho\chi} v_\rho^2 \right) },
\end{equation}
and eigenvalues:
\begin{align}
m_{S_{3{R/I}}}^2 &= \frac{1}{2} \left[ m_{\xi}^2 + m_{\chi}^2 + \sqrt{ \left(m_{\xi}^2 - m_{\chi}^2\right)^2 + 4 \lambda_{\xi \sigma \chi \rho v^2_\rho v^2_\sigma} } \right], \\
m_{S_{4{R/I}}}^2 &= \frac{1}{2} \left[ m_{\xi}^2 + m_{\chi}^2 - \sqrt{ \left(m_{\xi}^2 - m_{\chi}^2\right)^2 + 4 \lambda_{\xi \sigma \chi \rho v^2_\rho v^2_\sigma}} \right],
\end{align}
where $m_{\xi}^2 \equiv \mu_\xi^2 + \frac{1}{2} \lambda_{\phi\xi} v_\phi^2 + \frac{1}{2} \lambda_{\sigma\xi} v_\sigma^2 + \lambda_{\rho\xi} v_\rho^2,\quad
m_{\chi}^2 \equiv \mu_\chi^2 + \frac{1}{2} \lambda_{\phi\chi} v_\phi^2 + \frac{1}{2} \lambda_{\sigma\chi} v_\sigma^2 + \lambda_{\rho\chi} v_\rho^2$.
As before the masses will be degenerate between imaginary and real components and the mixing angle will have opposite sign ($\gamma_{S_{R}}=-\gamma_{S_{I}}$).

Lastly, the charged scalar $\eta^-$ does not mix and its mass is given by:
\begin{equation}\label{eq:m_eta}
m_{\eta}^2 = \mu_\eta^2 + \frac{1}{2} \lambda_{\phi\eta} v_\phi^2 + \frac{1}{2} \lambda_{\sigma\eta} v_\sigma^2 + \lambda_{\rho\eta} v_\rho^2.
\end{equation}

\subsection{Anomaly, stability and unitarity conditions}
With the field content and interactions described in Section \ref{Sec:model}, we have verified that our  model is anomaly free and checked the cancellation of both gauge and gravitational anomalies; the related conditions are listed in Appendix \ref{sec:anom.canc}. 
We recall that regarding the scalar potential, we choose $-\mu_\phi^2$, $-\mu_\sigma^2$, and $-\mu_\rho^2$ as negative and   a CP-conserving scenario:  with the trilinear coupling $A_{\phi\eta\varphi}$ and all quartic terms to be real. The other restrictions we impose on the parameters are due to  potential stability and unitarity conditions  that are detailed in Appendix~\ref{sec:stability-unitarity}.

\section{Neutrino mass generation }\label{Sec:mass-generation}
In this section we will explain the generation of the tiny active neutrino masses via an interplay of linear and inverse seesaw mechanisms (\cite{Abada:2018qok,Abada:2025edq}), here both dynamically generated  at two loop level. With the particle content and symmetries shown in Table~\ref{model}, we display the following complete neutrino Yukawa Lagrangian: 
\begin{eqnarray}
-\mathcal{L}_{Y}^{(\nu)} &=&
\sum_{i=1}^{3} (y_\nu)_i\,\overline{l_{iL}}\,\widetilde{\phi}\,\nu_R
\;+\; M_N\,\overline{\nu_R}\,N_R^{\,C}
\;+\; \sum_{i=1}^{3} (y_{\Psi})_{i1}\,\overline{l_{iL}}\,\eta\,\Psi_{R}
\;+\; y_N\,\overline{\Psi_L}\,\varphi\,N_R
\nonumber\\[2pt]
&&
+\; y_{\Omega}\,\overline{\Omega_L}\,\xi\,\Psi_{R}
\;+\; (z_{\Psi})_{1}\,N_R\,\varphi^{\ast}\,\overline{\Psi_{R}^{\,C}}
\;+\; m_{\Omega}\,\overline{\Omega_L}\,\Omega_R
\nonumber\\[2pt]
&&
+\; (z_{\Psi})_{2}\,\sigma\,\overline{\tilde{\Psi}_{R}^{\,C}}\,\tilde{\Psi}_{R}
\;+\; z_{\Omega}\,\overline{\Psi_L}\,\chi\,\Omega_R
\;+\; \text{h.c.}\, .
\label{eq:Yukawa_lagrangian}
\end{eqnarray}

In Section~\ref{Sec:scalar-potential}, we have already discussed how the scalar fields acquire their masses.
In the remainder of this section, we analyze the corresponding case for neutral leptons—namely, the neutrino and $\Psi$ field  sectors—while the $\Omega$ fields already obtain their masses at tree level, as can be directly inferred from Eq.~\eqref{eq:Yukawa_lagrangian}.

\subsection{Generation of the LISS mechanism}\label{sec:neutrino_mass}

\begin{figure}[h!]
    \centering
    \begin{tabular}{cc}
        \includegraphics[width=0.50\textwidth]{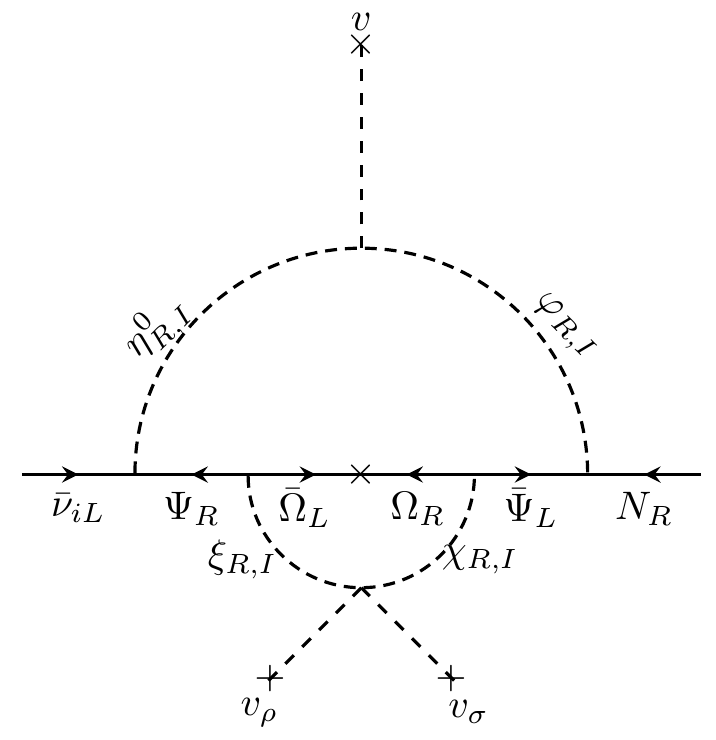} &
        \includegraphics[width=0.50\textwidth]{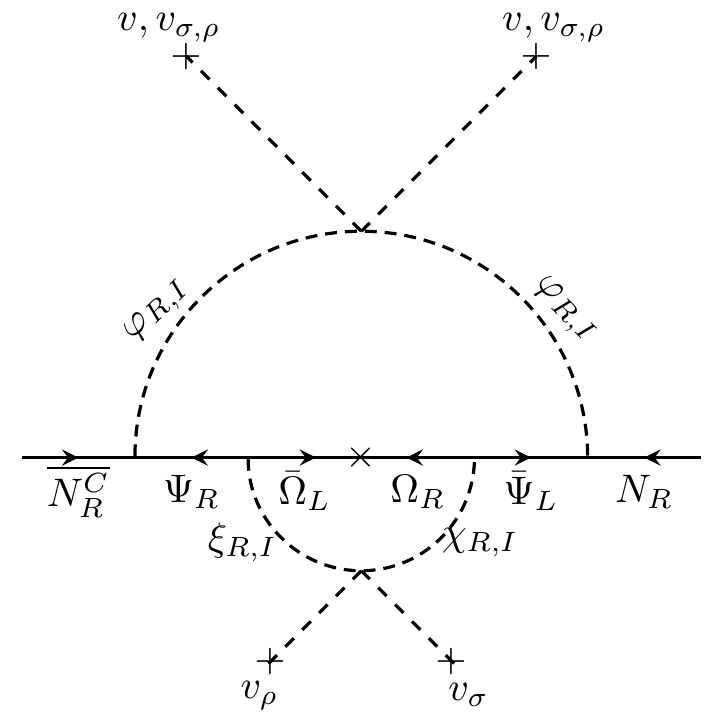} \\
        (a) & (b)
    \end{tabular}
    \caption{Two-loop diagrams for the neutrino mass submatrix $\varepsilon$ (a) and for the lepton number violating Majorana mass term $\mu$ (b).}
    \label{epdiagram}
\end{figure}

The neutrino Yukawa interactions of Eq.~(\ref{eq:Yukawa_lagrangian}) 
induce the following neutrino mass
terms:
\begin{align}
-\mathcal{L}_{\text{mass}}^{(\nu)} &=
\frac{1}{2}
\begin{pmatrix}
\overline{\nu_L^{C}} &
\overline{\nu_R} &
\overline{N_R}
\end{pmatrix}
M_{\nu}
\begin{pmatrix}
\nu_L \\
\nu_R^{C} \\
N_R^{C}
\end{pmatrix}
+\text{h.c.} \nonumber\\[2mm]
&= m_D\,\overline{\nu_L}\,\nu_R
+ \varepsilon\,\overline{\nu_L}\,N_R
+ M\,\overline{\nu_R^{C}}\,N_R
+ \tfrac{1}{2}\,\mu\,\overline{N_R^{C}}\,N_R
+\text{h.c.,}
\label{eq:Mass-Matr2}
\end{align}
where $m_{D{i}}\equiv(y_\nu)_i \frac{v}{\sqrt{2}}$ ($i=1,2,3$) is the Dirac mass column vector. With this model, the tiny active neutrino masses are produced by a combination of linear and
inverse seesaw mechanisms, where the full neutrino mass matrix expressed in
the basis $\left( \nu _{L},\nu _{R}^{C},N_{R}^{C}\right) $, obtains the
following structure: 
\begin{equation}
M_{\nu }=\left( 
\begin{array}{ccc}
0 & m_D & \varepsilon  \\ 
m_D^{T} & 0 & M \\ 
\varepsilon ^{T} & M & \mu 
\end{array}%
\right) ,  
\label{eq:Mnufull}
\end{equation}%
where $\nu _{{L}_i}$ ($i=1,2,3$) are the active neutrinos, whereas $\nu _{R}$ and $%
N_{R}$ are the sterile ones, 

Their lepton numbers are $L(\nu
_{L})=L(\nu _{R})=-L(N_{R})=1$. Consequently, the sources of lepton number
violation  are the $\varepsilon $ column vector and the $%
\mu $ entry. For the linear and inverse seesaw mechanisms to work
appropriately, thus assuming the seesaw condition,  the entries of the full neutrino mass matrix (\ref{eq:Mnufull})
must satisfy the following hierarchies: $\mu,\ \varepsilon_{i} \ll m_{D_{i}} \ll M$
 (for $%
i=1,2,3$). Furthermore, the column vector $m_D$ and the entry $M$ arise
at tree level, whereas the column vector $\varepsilon $ and the entry $\mu $ are
generated at two-loop level as indicated in the Feynman diagrams (a) and (b) of Figure %
\ref{epdiagram},
respectively and they are given by: 
\begin{eqnarray}
{m_{D}}_{i} &=&\left( y_{\nu }\right) _{i}\frac{v}{\sqrt{2}},\hspace{0.7cm}\hspace{%
0.7cm}i=1,2,3\ , \\
\varepsilon _{i} &=&\sum_{s=1}^{2}\sum_{p=1}^{2}\frac{\left(
y_{\Psi }\right) _{i1} y_{\Omega }y_{N}z_{\Omega
}m_{\Omega }}{4(4\pi )^{4}}  \notag \\
&&\times \int_{0}^{1}d\alpha \int_{0}^{1-\alpha }d\beta \frac{1}{\alpha
(1-\alpha )}\left[ \left( R_{S_{R}}\right) _{1s}\left( R_{S_{R}}\right)
_{2s}\left( \widetilde{R}_{S_{R}}\right) _{1p}\left( \widetilde{R}%
_{S_{R}}\right) _{2p}I\left( m_{\Omega
_{k}}^{2},m_{R_{s}R_{(p+2)}}^{2},m_{R_{s}I_{(p+2)}}^{2}\right) \right.   \notag \\
&&+\left. \left( R_{S_{I}}\right) _{1s}\left( R_{S_{I}}\right) _{2s}\left( 
\widetilde{R}_{S_{I}}\right) _{1p}\left( \widetilde{R}_{S_{I}}\right)
_{2p}I\left( m_{\Omega
_{k}}^{2},m_{I_{s}R_{(p+2)}}^{2},m_{I_{s}I_{(p+2)}}^{2}\right) \right], \hspace{0.3cm}\hspace{%
0.7cm}i=1,2,3\ , \label{Eq:epsilonterm} \\
\mu  &=&\sum_{s=1}^{2}\sum_{p=1}^{2}\frac{\left( z_{\Psi
}\right)_{1}y_{\Omega }z_{\Omega }y_{N}m_{\Omega }}{%
4(4\pi )^{4}}  \notag \\
&&\times \int_{0}^{1}d\alpha \int_{0}^{1-\alpha }d\beta \frac{1}{\alpha
(1-\alpha )}\left[ \left( R_{S_{R}}\right) _{2s}\left( R_{S_{R}}\right)
_{2s}\left( \widetilde{R}_{S_{R}}\right) _{1p}\left( \widetilde{R}%
_{S_{R}}\right) _{2p}I\left( m_{\Omega
_{k}}^{2},m_{R_{s}R_{(p+2)}}^{2},m_{R_{s}I_{(p+2)}}^{2}\right) \right.   \notag \\
&&-\left. \left( R_{S_{I}}\right) _{2s}\left( R_{S_{I}}\right) _{2s}\left( 
\widetilde{R}_{S_{I}}\right) _{1p}\left( \widetilde{R}_{S_{I}}\right)
_{2p}I\left( m_{\Omega _{k}}^{2},m_{I_{s}R_{(p+2)}}^{2},m_{I_{s}I_{(p+2)}}^{2}\right) \right]
\ . \label{Eq:muterm}
\end{eqnarray}%
The two-loop integral entering in Eqs. (\ref{Eq:epsilonterm}) and (\ref{Eq:muterm}) takes the form \cite{Kajiyama:2013rla,Hernandez:2021xet}: 
\begin{eqnarray}
I(m_{1}^{2},m_{2}^{2},m_{3}^{2}) 
&=& \frac{
m_{1}^{2}m_{2}^{2}\,\ln\!\left(\tfrac{m_{2}^{2}}{m_{1}^{2}}\right)
+ m_{2}^{2}m_{3}^{2}\,\ln\!\left(\tfrac{m_{3}^{2}}{m_{2}^{2}}\right)
+ m_{3}^{2}m_{1}^{2}\,\ln\!\left(\tfrac{m_{1}^{2}}{m_{3}^{2}}\right)
}{
(m_{1}^{2}-m_{2}^{2})(m_{1}^{2}-m_{3}^{2})
}\ , \notag \\[2mm]
m_{a_{k}b_{l}}^{2} 
&=& \frac{\beta\, m_{(S_{k})_{a}}^{2} + \alpha\, m_{(S_{l})_{b}}^{2}}{\alpha (1-\alpha)},
\qquad k,l=1,2,\quad a,b \in \{R,I\}\ .
\end{eqnarray}

Finally, $R_{S_{R/I}}$ and $\widetilde{R}_{S_{R/I}}$ in Eqs. (3.5) and (3.6) represent the rotation matrices that relate the interaction states to the physical CP-even and CP-odd scalar fields, as defined in Eqs.~\eqref{eq:R} and~\eqref{eq:R_tilde}. Note that these four rotation matrices must fulfill the following orthogonality conditions:

\begin{equation}
\sum_{s=1}^{2} (R_{S_a})_{1s}\,(R_{S_a})_{2s} = 0 
\qquad ; \quad
\sum_{s=1}^{2} (\widetilde{R}_{S_a})_{1s}\,(\widetilde{R}_{S_a})_{2s} = 0,
\qquad a \in \{R,I\}\, .
\end{equation}\label{eq:orthogonal_relations}
These conditions will be instrumental in understanding the discussion of the results presented in Section~\ref{sec:results}.


In the following, we proceed with the determination of the neutrino mass spectrum. Starting from the full neutrino mass matrix 
$M_{\nu }$
in Eq.~\eqref{eq:Mnufull}, we first perform a perturbative block diagonalization under the seesaw conditions $\mu,m_D,\varepsilon \ll M$: 
\begin{equation}
U^{\prime T} M_{\nu \ }
U^{\prime} \simeq \begin{pmatrix}
\mathcal{M^{\text{light}}_{\nu}} & 0 \\
0 & \mathcal{M_H}
\end{pmatrix}\ , 
\end{equation}
where $\mathcal{M^{\text{light}}_{\nu}}$ is the $3\times 3$ mass matrix of the light (active) neutrinos  and 
$\mathcal{M_H}$ is the $2\times 2$ mass matrix of the heavy neutral fermions. 
This transformation is achieved through the matrix $U'$, which, in the seesaw limit, takes the form
\begin{equation}
U^\prime \simeq
 \begin{pmatrix}
1 - \frac{1}{2} RR^{\dagger} & R \\
-R^{\dagger} & 1 - \frac{1}{2} R^{\dagger} R
\end{pmatrix}, \quad \text{with} \quad R \equiv \Big(m_D\quad \varepsilon\Big) \mathcal{M}_H^{-1}.
\end{equation}
Explicitly, the matrix $R$ is given by:
\begin{equation}
R = \begin{pmatrix}
\frac{-\mu m_D+M \varepsilon}{M^2} & \frac{m_D}{M}
\end{pmatrix}\ .\label{eq:rmatrix}
\end{equation}
Due to the presence of the non-vanishing column LNV vector $\varepsilon$ and the LNV Majorana mass parameter $\mu$, small masses for the active neutrinos are generated within the scotogenic LISS framework. The light neutrinos thus acquire masses from the combined action of the linear~\cite{Akhmedov:1995ip,Akhmedov:1995vm,Malinsky:2005bi} and inverse seesaw mechanisms~\cite{Wyler:1982dd, Mohapatra:1986bd, GonzalezGarcia:1988rw, Akhmedov:1995ip, Akhmedov:1995vm, Malinsky:2009df, Abada:2014vea}, both dynamically generated in the present model. This yields the following light-neutrino mass matrix:
\begin{equation}
 \mathcal{M^{\text{light}}_{\nu}}\simeq
   \frac{\mu}{M^2} m_D m_D^T-\frac{1}{M}\left(\varepsilon m_D^T+m_D \varepsilon^T\right)\ ,
\end{equation}
whereas the heavy-sector mass matrix $\mathcal{M}_H$ is approximately given by:
\begin{equation}
    \mathcal{M}_H\simeq
\begin{pmatrix}
0 & M \\
M & \mu
\end{pmatrix}\ .
\end{equation}

A final diagonalization step is performed on these two blocks. The light-neutrino mass matrix $\mathcal{M^{\text{light}}_{\nu}}$ 
 is diagonalized by a unitary matrix $R_{\nu}$,
while $\mathcal{M}_H$ is diagonalized by a unitary matrix $R_M$, yielding the physical neutrino masses. In this model, the light-neutrino sector contains one massless state, with the remaining two masses constrained by the solar and atmospheric mass-squared differences observed in oscillation experiments.
The heavy neutrino mass spectrum, determined by $\mathcal{M}_H$ and associated with the LNV parameter $\mu$ related to the inverse seesaw component of the full mass matrix $M_{\nu}$ in Eq.~\eqref{eq:Mnufull}, consists of a pair of quasi-Dirac heavy neutrinos~\cite{Valle:1982yw,Anamiati:2016uxp,Arbelaez:2021chf}. Their approximate masses are:
\begin{align} 
M_{N^{-}} &\simeq
 -M+\frac{\mu }{2}, \\
M_{N^{+}} &\simeq
 M+\frac{\mu }{2}.
\end{align}
Finally, the full neutrino mixing matrix $U$ is then given by:
\begin{equation}
U \simeq
\begin{pmatrix}
(1 - \frac{1}{2} RR^{\dagger})R_{\nu} & RR_{M} \\
-R^{\dagger}R_{\nu} & (1 - \frac{1}{2} R^{\dagger} R)R_{M}
\end{pmatrix}.
\end{equation}

\subsection{Majorana mass generation for the dark neutral leptons 
$\Psi_{R}$ and $\tilde{\Psi}_{R}$ }\label{sec:PsiMass}
The fermions $\Psi_{R}$ and $\tilde{\Psi}_{R}$ behave differently under the symmetries of the model, which leads to qualitatively distinct origins for their Majorana masses. The $\tilde{\Psi}_{R}$ acquires a Majorana mass term from Eq.~\eqref{eq:Yukawa_lagrangian}, after $U(1)'$ breaking given by: $\langle\sigma\rangle=v_\sigma/\sqrt{2}$,
\begin{equation}
m_{\tilde{\Psi}_{R}}=\frac{(z_\Psi)_2\,v_\sigma}{\sqrt{2}}\;.
\end{equation}
On the other hand, a Majorana mass for $\Psi_{R}$ is forbidden at tree level by the $\mathbb{Z}_3$ symmetry, but arises radiatively at one loop through the internal loop of the diagrams of Fig~\ref{epdiagram}. This Majorana mass is given by:

\begin{align}\label{eq:m_psi}
m_{\Psi_{R}}
&=\frac{z_{\Omega}\,y_{\Omega}\,m_{\Omega}}{8\pi^{2}}
\sum_{p=1}^{2} \Bigg[
(\widetilde{R}_{S_{R}})_{2p}(\widetilde{R}_{S_{R}})_{1p}\,
\frac{m^{2}_{S_{(p+2)R}}}{m^{2}_{S_{(p+2)R}}-m_{\Omega}^{2}}
\ln\!\left(\frac{m^{2}_{S_{(p+2)R}}}{m_{\Omega}^{2}}\right) \notag \\[2mm]
&\hspace{2cm}
-(\widetilde{R}_{S_{I}})_{2p}(\widetilde{R}_{S_{I}})_{1p}\,
\frac{m^{2}_{S_{(p+2)I}}}{m^{2}_{S_{(p+2)I}}-m_{\Omega}^{2}}
\ln\!\left(\frac{m^{2}_{S_{(p+2)I}}}{m_{\Omega}^{2}}\right)
\Bigg]\!.
\end{align}

\section{Phenomenology} \label{sec:pheno}

Our model provides a rich phenomenology. In this section, we focus on laboratory probes: we briefly discuss how collider searches could test our framework and outline future directions to explore, while primarily concentrating on charged-lepton–flavor-violating (cLFV) observables and on the predictions for Muonium decays and Muonium–antimuonium oscillations. Cosmological aspects, such as dark-matter production and domain-wall dynamics, are left for future work.

\subsection{Collider searches}\label{sec:colliders}

Here we provide a concise qualitative discussion of the collider signatures of Dirac sterile neutrinos  $\Omega_{L,R}$, heavy $Z^\prime$, and dark matter candidates.
The heavy quasi-Dirac sterile neutrinos can be produced at the LHC in association with a SM charged lepton via quark–antiquark annihilation, which corresponds to the Drell–Yan (DY) mechanism. Furthermore, like the $U(1)^\prime$ scenario in Ref.~\cite{Deppisch:2013cya}, our model predicts two-body sterile neutrino decays $N \rightarrow l^\pm_i W^\mp$, $\nu_i Z$ and $\nu_i h$ (with flavor index $i = 1,2,3$), suppressed by the small active–sterile mixing angle $\theta \sim \mathcal{O}(10^{-3})$, which should be constrained to maintain charged LFV decays well below present experimental limits and remain consistent with unitarity requirements~\cite{Abada:2018nio,Fernandez-Martinez:2016lgt}.
These decays mediate three-body final states such as $N \rightarrow l^+_i l^-_j \nu_k$, $N \rightarrow l^-_i u_j \bar{d}_k$, $N \rightarrow b\bar{b}\nu_k$ (where $i,j,k = 1,2,3$ are flavor indices), which are also present in Ref.~\cite{Deppisch:2013cya}. Consequently, we expect that our total cross-section predictions for the LFV process $pp \rightarrow NN \rightarrow e^\pm \mu^\mp 4j$ and sterile neutrino lifetimes will align closely with those of Ref.~\cite{Deppisch:2013cya}. However, a slight deviation is expected in the decay rate of $N \rightarrow l^{+}_i l^{-}_j \nu_k$, as our model includes contributions from off-shell $W$ gauge bosons.

Regarding $Z^\prime$ production, it is worth noting that the $Z^\prime$ can be produced at proton–proton colliders through both the DY mechanism and Vector Boson Fusion (VBF). Once produced, the heavy $Z^\prime$ gauge boson can decay into fermion–antifermion pairs, leading to characteristic signatures with dilepton or dijet final states, such as
$pp \rightarrow Z^\prime \rightarrow l^{+}l^{-}$,
$pp \rightarrow Z^\prime \rightarrow 2j$,
$pp \rightarrow Z^\prime, 2j \rightarrow l^{+}l^{-} 2j$, and
$pp \rightarrow Z^\prime, 2j \rightarrow 4j$.
Detailed studies of $Z^\prime$ production at colliders within $B-L$ gauge theories are presented in Ref.~\cite{Hernandez:2021uxx}. The constraint $\frac{M_{Z^{\prime}}}{g_X} > 7~\text{TeV}$, derived from LEP I/II measurements of $e^{+}e^{-}\rightarrow l^{+}l^{-}$~\cite{LEP:2004xhf,Carena:2004xs,Das:2021esm}, is further supported by LHC searches~\cite{ATLAS:2019erb,CMS:2021ctt}. Additional bounds on $\frac{M_{Z^{\prime}}}{g_{B-L}}$ from LEP II and projected sensitivities at future $e^{+}e^{-}$ colliders (e.g., ILC) are analyzed in Ref.~\cite{Das:2021esm}.

On the other hand, thanks to the conserved $\widetilde{\mathbb{Z}}_2$ and $\mathbb{Z}_3$ symmetries, the dark matter candidates are pair-produced. The neutral components of the inert doublet can be pair-produced through two primary mechanisms: DY processes mediated by the $Z$ boson or VBF. At colliders, these production channels lead to a distinctive signature of two jets accompanied by missing transverse energy. The DY production channel of inert doublet pairs at the LHC is analyzed in~\cite{CarcamoHernandez:2023atk}. In addition, for the VBF production channel specifically, comprehensive analyses of the resulting collider signatures are presented in Ref.~\cite{Dutta:2017lny}. Furthermore, as in the model of Ref.~\cite{CarcamoHernandez:2023atk}, in the fermionic dark matter scenario, the pair production of the charged inert doublet components via the Drell–Yan process, followed by their subsequent decays, can generate a distinctive opposite-sign dilepton plus missing transverse energy (MET) signature. An observed excess of such events above the Standard Model background predictions at the LHC would provide supporting evidence for this model. While a comprehensive analysis of these collider signatures represents an important future direction, it falls outside the scope of the current study.

Finally, the ratio $\frac{M_{Z^{\prime}}}{g_X}$ can also be constrained from dark matter direct detection measurements on the upper limit of the spin-independent WIMP–nucleon cross section. In the fermionic dark matter scenario, $Z^{\prime}$ exchange leads to spin-independent scattering with nucleons, along similar lines to Ref.~\cite{Abada:2021yot}. In the limit of small momentum transfer, the $Z'$ exchange gives rise to a spin-independent scattering with nucleons.
The corresponding cross section per nucleon takes the form~\cite{Alves:2015pea}
\begin{equation}
    \sigma \simeq \frac{g_{X}^4}{\pi} \frac{m_n^2\, m_{\Psi_R}^2}{(m_n + m_{\Psi_R})^2\, M_{Z'}^4}\,,
\end{equation}
where $m_n$ is the nucleon mass, and $f_n \sim 0.3$ is the hadron matrix element. Given that the mass for our fermionic dark matter candidate $\Psi_R$ arises at one loop, we consider its mass to be around the $10$ GeV scale, which implies that its spin independent cross section for direct detection should be smaller than $4.8\times 10^{-47}$cm$^2$, as follows from the data provided by the LUX experiment~\cite{LZ:2024zvo}. Consequently, the constraints arising from the LUX experiment~\cite{LZ:2024zvo} set the ratio $\frac{M_{Z^{\prime}}}{g_X}$ to be larger than about $40~\text{TeV}$ for a $10~\text{GeV}$ fermionic dark matter mass.

\subsection{Charged lepton flavor violation} \label{sec:clfv}
Here we discuss the implications of our model in charged lepton flavor violating decays such as $\ell_\alpha\to \ell_\beta\ell_\beta\ell_\rho$, $\text{CR}(\mu -e, \text{ N})$, and the radiative ones, $\ell\to \ell'\gamma$.

In this model, the neutral spectrum  is composed of seven  heavy sterile neutrinos and among them, given the $\mathbb{Z}_3$ charge assignment only $\nu_R$ and 
$\psi_{1R}$ provide  contribution to  the penguin and box diagrams of these cLFV observables. The scalar sector may also contribute and due to the symmetries, only the charged component of the scalar doublet $\eta$ provides a contribution at one-loop level.
\subsubsection{Radiative cLFV decays}
As discussed above, the computation of the decay amplitude for radiative $l_i \to l_j \gamma$ ($\mu \to e\gamma$, $\tau \to \mu \gamma$ and $\tau \to e\gamma $) decays due to both the neutral fermions and the scalar field $\eta$ contributions~\cite{Langacker:1988up, Lavoura:2003xp, Hue:2017lak, CarcamoHernandez:2020pnh, Bonilla:2023egs, Bonilla:2023wok, CarcamoHernandez:2023atk, Batra:2023mds, Ma:2001mr, Toma:2013zsa, Vicente:2014wga, Lindner:2016bgg, Abada:2022dvm} leads to the following branching ratio: 
\begin{align}
\text{Br}\left( l_{i}\rightarrow l_{j}\gamma \right) & =\left\vert
\left( 
\frac{\alpha _{W}^{3}s_{W}^{2}m_{l_{i}}^{5}}{256\pi ^{2}m_{W}^{4}\Gamma _{i}}%
\right)^{1/2}G_{ij}+
\left( \frac{3\left( 4\pi \right)^{3}\alpha _{EM}}{4G_{F}^{2}}\right)^{1/2}\sum_{k=1}^{3}\frac{z_{kj}^{\ast }z_{ki}}{2\left(
4\pi \right) ^{2}}\frac{1}{m_{\eta ^{\pm}}^{2}}F_{2}\left( \xi \right)
\right\vert ^{2}, \label{eq:br.radiative}\\
G_{ij}& \simeq \sum_{k=1}^{3}\left( \left[ \left( 1-\frac{1}{2}RR^{\dagger }\right)
R_{\nu }\right] ^{\ast }\right) _{ik}\left( \left( 1-\frac{1}{2}RR^{\dagger }\right)
R_{\nu }\right) _{jk}G_{\gamma }\left( \frac{m_{\nu _{k}}^{2}}{m_{W}^{2}}%
\right)   \notag \\
& \qquad +2\sum_{k=1}^{2}\left( R\right)^{\ast }_{ik}\left( R\right)
_{jk}G_{\gamma }\left( \frac{m_{N_{k}}^{2}}{m_{W}^{2}}\right) , \\
G_{\gamma }(x)& \equiv \frac{10-43x+78x^{2}-49x^{3}+18x^{3}\ln x+4x^{4}}{%
12\left( 1-x\right) ^{4}}, \\
F_{2}\left( x\right) & =\frac{1-6x+3x^{2}+2x^{3}-6x^{2}\ln x}{6\left(
1-x\right) ^{4}}\ .\label{eq:F2}
\end{align}

\noindent Here 
$z_{is} = \sum_{k=1}^3 y_\Psi^{ks}\,(V_{lL}^\dagger)_{ik}$, 
 $V_{lL}$ being the charged lepton mixing matrix in general, 
$R$ and $R_{\nu}$ are defined in Section~\ref{sec:neutrino_mass}, 
 $s_W$ corresponds to the sine of the weak mixing angle and $\Gamma_i$ is the total decay width of the charged lepton with its mass denoted by $m_{l_{i}}$. For simplicity, in our calculation we work in the charged–lepton mass basis, 
so that $(V_{lL}^\dagger)_{ik} = \delta_{ik}$ and $z_{is}$ reduces to the Yukawa couplings $y_\Psi^{is}$. 

The first term in Eq.~\eqref{eq:br.radiative} corresponds to the contribution of the neutral fermions, while the second one arises from the charged component of the dark scalar doublet $\eta$. The argument of the associated loop function is
$\xi = \tfrac{m_{\Psi_R}^2}{m_{\eta^{\pm}}^2}$,
which is given by the mass ratio of $\Psi_R$ and $\eta^{\pm}$, as defined in \eqref{eq:m_psi} and \eqref{eq:m_eta}, respectively.

\subsubsection{ $\text{CR}(\mu -e, \text{N})$}
The $\mu - e$ conversion occurs in a muonic atom formed when a muon is captured, falling into the first state of a target nucleus $N$. The conversion rate is defined as
\begin{equation} \label{eq:CR:def}
    \text{CR}(\mu -e, \text{ N}) \equiv \frac{\Gamma (\mu^- + N \to e^- +N)}{\Gamma (\mu^- + N \to \text{ all)}}\,.
\end{equation}
The box and penguin diagrams inducing $\mu - e$ conversion receive both neutrino and scalar contributions which are given by:
\begin{eqnarray}
\text{CR}(\mu -e,\text{N}) &=&\left\vert \sqrt{\frac{2\,G_{F}^{2}\,\alpha
_{W}^{2}\,m_{\mu }^{5}}{(4\pi )^{2}\,\Gamma _{\text{capt}}(Z)}\left[
4\,V^{(p)}\left( 2\,\tilde{F}_{u}^{\mu e}+\tilde{F}_{d}^{\mu e}\right)
+4\,V^{(n)}\left( \tilde{F}_{u}^{\mu e}+2\,\tilde{F}_{d}^{\mu e}\right)
+D\,G_{\gamma }^{\mu e}\frac{s_{W}^{2}}{2\sqrt{4\pi \alpha }}\right] }%
\right. \notag \\
&&+\left. \left[ \frac{p_{e}E_{e}m_{\mu }^{3}G_{F}^{2}\alpha _{\text{EM}%
}^{3}Z_{\text{eff}}^{4}F_{p}^{2}}{8\pi ^{2}\,Z\,\Gamma _{\text{capt}}}\left[
\left\vert (Z+N)\left(g_{LS}^{(0)}\right) +(Z-N)\left(
g_{LS}^{(1)}\right) \right\vert ^{2}\right. \right. \right.\notag  \\
&&+\left. \left. \left. \left\vert (Z+N)\left(
g_{RS}^{(0)}\right) +(Z-N)\left(
g_{RS}^{(1)}\right) \right\vert ^{2}\right] \right] ^{\frac{1}{2%
}}\right\vert ^{2},
\end{eqnarray}
where $G_F$ is the Fermi constant, $m_\mu$ the muon mass, $\alpha_{\text{EM}} \equiv e^2/(4\pi)$, with $s_W$ corresponding to the sine of the weak mixing angle and $\Gamma_\text{capt}(Z)$ denotes the capture rate of a nucleus with atomic number $Z$~\cite{Kitano:2002mt}. The form factors $\tilde{F}_{q}^{\mu e}$ ($q=u,d$) are given by 
\begin{equation}
    \tilde F_q^{\mu e} = Q_q \, s_W^2 F^{\mu e}_\gamma+F^{\mu e}_Z \left(\frac{{I}^3_q}{2}-Q_q\, s_W^2\right) + \frac14 F^{\mu eqq}_\text{box}\,,
    \label{eq:tildeFqmue}
\end{equation}
where $Q_q$ denotes the quark electric charge ($Q_u=2/3$, $Q_d=-1/3$) and ${I}^3_q$ is the weak isospin ($\mathcal{I}^3_u=1/2$, $\mathcal{I}^3_d=-1/2$). The quantities $F^{\mu e}_\gamma$, $F^{\mu e}_Z$ and $F^{\mu eqq}_\text{box}$ correspond to the different form factors of the diagrams, and $G^{\mu e}_\gamma$ corresponds to the dipole term; all expressions are collected in Appendix~\ref{app:C}. The relevant information (nuclear form factors and averages over the atomic electric field) is encoded in the form factors $F_p$, $D$, $V^{(p)}$, and $V^{(n)}$. In our analysis, we use the numerical values presented in Ref.~\cite{Kitano:2002mt}.

In the above, the effective nucleon couplings $g_{LS}^{(0)}$ and $g_{LS}^{(1)}$ can be expressed in terms of the quark--level scalar couplings as
\begin{equation}\label{eq:nucleon_effective_couplings}
\begin{aligned}
g_{LS}^{(0)} &= \frac{1}{2} \sum_{q = u,d,s} 
   \bigl( g_{LS}^{(q)}\, G_S^{(q,p)} + g_{LS}^{(q)}\, G_S^{(q,n)} \bigr), \\
g_{LS}^{(1)} &= \frac{1}{2} \sum_{q = u,d,s} 
   \bigl( g_{LS}^{(q)}\, G_S^{(q,p)} - g_{LS}^{(q)}\, G_S^{(q,n)} \bigr).
\end{aligned}
\end{equation}
Only the light quark flavors $u$, $d$, and $s$ provide sizable contributions. In Appendix~\ref{app:LFV-Yukawa}, we express the quark-level scalar couplings $g_{\text{LS}}^{(q)}$ in terms of the parameters of our model.
The corresponding scalar form factors entering the nucleon matrix elements are taken from Refs.~\cite{Kosmas:2001mv,Kuno:1999jp}.
\begin{equation}
\begin{aligned}
& G_S^{(u, p)} = G_S^{(d, n)} = 5.1, \qquad 
  G_S^{(d, p)} = G_S^{(u, n)} = 4.3, \\
& G_S^{(s, p)} = G_S^{(s, n)} = 2.5.
\end{aligned}
\end{equation}

\subsubsection{$\ell_\alpha\to \ell_\beta\ell_\beta\ell_\rho$}
The branching ratio for the three body charged lepton flavor violating decay $\ell _{\beta }\rightarrow \ell _{\alpha }\overline{\ell _{\alpha
}}\ell _{\alpha }$ takes the form \cite{Ilakovac:1994kj,Alonso:2012ji,Toma:2013zsa,Vicente:2014wga}: 
\begin{align}\label{eq:Br_lepton_violating_decay}
\mathrm{BR}(\ell _{\beta }\rightarrow \ell _{\alpha }\overline{\ell _{\alpha
}}\ell _{\alpha })=& \left\vert \left[ \frac{\alpha _{W}^{4}}{24576\,\pi ^{3}%
}\,\frac{m_{\beta }^{4}}{M_{W}^{4}}\,\frac{m_{\beta }}{\Gamma _{\beta }}%
\times \left\{ 2\left\vert \frac{1}{2}F_{\text{box}}^{\beta 3\alpha
}+F_{Z}^{\beta \alpha }-2s_{W}^{2}\,(F_{Z}^{\beta \alpha }-F_{\gamma
}^{\beta \alpha })\right\vert ^{2}\right. \right. \right.   \notag \\
& +\left. \left. \left. 4s_{W}^{4}\,|F_{Z}^{\beta \alpha }-F_{\gamma
}^{\beta \alpha }|^{2}+16s_{W}^{2}\,\mathrm{Re}\left[ (F_{Z}^{\beta \alpha }-%
\frac{1}{2}F_{\text{box}}^{\beta 3\alpha })\,G_{\gamma }^{\beta \alpha \ast }%
\right] \right. \right. \right.   \notag \\
& -\left. \left. 48s_{W}^{4}\,\mathrm{Re}\left[ (F_{Z}^{\beta \alpha
}-F_{\gamma }^{\beta \alpha })\,G_{\gamma }^{\beta \alpha \ast }\right]
+32s_{W}^{4}\,|G_{\gamma }^{\beta \alpha }|^{2}\left[ \log \frac{m_{\beta
}^{2}}{m_{\alpha }^{2}}-\frac{11}{4}\right] \right] ^{\frac{1}{2}}\right.  
\notag \\
& +\left. \left[ \frac{3(4\pi )^{2}\alpha _{\mathrm{EM}}^{2}}{8\,G_{F}^{2}}%
\left[ |A_{ND}|^{2}+|A_{D}|^{2}\left( \frac{16}{3}\log \left( \frac{%
m_{_{\beta }}}{m_{\alpha }}\right) -\frac{22}{3}\right) +\frac{1}{6}%
|B|^{2}\right. \right. \right.   \notag \\
& +\left. \left. \left. \left( -2A_{ND}\,A_{D}^{\ast }+\frac{1}{3}%
A_{ND}B^{\ast }-\frac{2}{3}A_{D}B^{\ast }+\mathrm{h.c.}\right) \right] \text{%
BR}\left( \ell _{\beta }\rightarrow \ell _{\alpha }\nu _{\beta }\overline{%
\nu _{\alpha }}\right) \right] ^{\frac{1}{2}}\right\vert ^{2}.
\end{align}

The dipole form factor is given by 

\begin{equation}\label{eq:AD}
    A_D=\sum_{i=1}^3 \frac{z_{i \beta}^* z_{i \alpha}}{2(4 \pi)^2} \frac{1}{m_{\eta^{+}}^2} F_2\left(\xi\right).
\end{equation}
The form factor $A_{ND}$ is generated from the non-dipole photon penguin diagrams and is expressed as
\begin{equation} \label{eq:AND}
    A_{ND}=\sum_{i=1}^3\frac{z_{i\beta}^*\, z_{i\alpha}}{6 (4 \pi)^2} \frac{1}{m_{\eta^+}^2} G_2\left(\xi\right).
\end{equation}
In addition to that, the contribution of box diagrams generates the form factor $B$, which is given by
\begin{equation} \label{eq:B}
    e^2 B = \frac{1}{(4 \pi)^2 m_{\eta^+}^2} \sum_{i,\, j = 1}^3 \left[\frac12 D_1(\xi_i,\, \xi_j)\, z_{j \beta}^*\, z_{j \beta}\, z_{i \beta}^*\, z_{i \alpha} + \sqrt{\xi_i\, \xi_j} D_2(\xi_i,\, \xi_j)\, z_{j \beta}^*\, z_{j \beta}^*\, z_{i \beta}\, z_{i \alpha}\right],
\end{equation}
where the different loop functions take the form
\begin{align}
    G_2(x) &= \frac{2 - 9 x + 18 x^2 - 11 x^3 + 6 x^3 \log x}{6 (1-x)^4}\,, \\
    D_1(x,y) &= -\frac{1}{(1-x) (1-y)} - \frac{x^2 \log x}{(1-x)^2 (x-y)} - \frac{y^2 \log y}{(1-y)^2 (y-x)}\,, \\
    D_2(x,y) &= -\frac{1}{(1-x) (1-y)} - \frac{x \log x}{(1-x)^2 (x-y)} - \frac{y \log y}{(1-y)^2 (y-x)}\,.
\end{align}
and $F_2(x)$ was already defined in Eq.~\eqref{eq:F2}. The remaining form factors present in the neutrino contribution — also relevant for $\mu-e$ conversion — are collected in Appendix~\ref{app:C}.

\subsection{Muonium decays and Muonium-antimuonium oscillation}\label{sec:muonium}

Formed in matter when a $\mu^+$ slows down and captures an $e^-$, Muonium (Mu) is a hydrogen-like bound state ($e^-\mu^+$)~\cite{Pontecorvo:1957cp}. 
Being free from hadronic interactions and with its electromagnetic structure precisely predicted in the SM, Mu provides an excellent system for high-precision measurements of fundamental constants and for testing new physics. 
Moreover, its cLFV transitions are essentially free from nuclear uncertainties.
In this work, we consider the contributions of our model to the cLFV decay $\text{Mu}\to e^+ e^-$~\cite{Cvetic:2006yg} and to Muonium–antimuonium conversion~\cite{Feinberg:1961zza}.
\bigskip

The decay $\text{Mu} \to e^+ e^-$ is strictly forbidden in the Standard Model.
In some of its extensions with right-handed neutrinos, this decay has been studied both at low energies and in the context of a future muon collider~\cite{Cvetic:2006yg}.
The cLFV Muonium decay rate can be written as
\begin{equation}\label{eq:Mudecay:BR}
\text{BR}(\text{Mu} \to e^+ e^-)  = 
\frac{\alpha_{\text{EM}}^3}{32 \pi^2\Gamma_\mu}
\frac{m_e^2 m_\mu^2}{(m_e + m_\mu)^3}
\sqrt{1 - 4\frac{m_e^2}{(m_e + m_\mu)^2}}
|\mathcal{M}_{\rm tot}|^2,
\end{equation}
where the amplitude $\mathcal{M}_{\rm tot}$ receives contributions from penguin and box diagrams with heavy neutrinos and dark scalar exchange~\cite{Cvetic:2006yg,Fukuyama:2021iyw}.
The strongest limit to date, set by the PSI experiment~\cite{Willmann:1998gd}, is
\begin{equation}
\text{BR}(\text{Mu} \to e^+ e^-) < 5.7 \times 10^{-6}.
\end{equation}
with the feasible future sensitivity lying around $\text{BR}(\text{Mu} \to e^+ e^-) < 1 \times 10^{-12}$~\cite{Gninenko:2012nt}.

\bigskip

While the decay $\text{Mu} \to e^+ e^-$ probes cLFV through a short-distance annihilation process, Muonium–antimuonium oscillations test a qualitatively different regime.
Instead of a decay within the bound state, the system undergoes a LFV transition into its antiparticle, $\overline{\text{Mu}} = e^{+}\mu^{-}$. This conversion is a striking signature of new physics~\cite{Feinberg:1961zza}: its observation would constitute unambiguous evidence for $|\Delta L_{e,\mu}|=2$ interactions and is sensitive to operators that are not necessarily the same as those contributing to the cLFV decay.

This transition can be described by the effective interaction for
$e^- \mu^+ \to e^+ \mu^-$, which for a $(V-A)\times(V-A)$ structure~\cite{Kuno:1999jp} reads
\begin{equation}\label{eq:Leff_muonium} \mathcal{L}^{\rm M\overline{M}}_\text{eff} = \frac{G_{\rm M\overline{M}}}{\sqrt{2}} \left(\overline{\mu}\gamma^\alpha (1 - \gamma_5)e\right) \left(\overline{\mu}\gamma_\alpha (1 - \gamma_5)e\right)\!, \end{equation}
where $G_{\rm M\overline{M}}$ parametrizes the strength of the new interaction.
It induces a small energy splitting between the Mu and $\overline{\text{Mu}}$ ground states,
\begin{equation}\label{eq:mu-barmu:Esplit} \delta_E^{\text{M}-\overline{\text{M}}} = \frac{8 \, G_F}{\sqrt{2} \, n^2 \pi a_0^3} \left(\frac{G_{\rm M\overline{M}}}{G_F}\right)\!, \end{equation}
where $n$ is the principal quantum number and $a_0$ the Bohr radius.

At present, no positive signal has been found; the best limit has been set at PSI~\cite{Willmann:1998gd}, where 
Muonium atoms are formed by electron capture when positively charged 
muons (from a very intense muon beam) are stopped in a SiO$_2$ powder
target and are given by~\cite{Willmann:1998gd}:
\begin{equation}
P(\text{Mu} \to \overline{\text{Mu}}) < 8.3 \times 10^{-11}
\quad \Rightarrow \quad
|G_{\rm M\overline{M}}| < 3.0 \times 10^{-3} G_F.
\end{equation}
Future experiments such as MACE (CSNS)~\cite{Snowmass:MACE2021} and J-PARC~\cite{Kawamura:2021lqk} aim to reach
$P(\text{Mu} \to \overline{\text{Mu}}) \sim 10^{-14}$.

\section{Results}\label{sec:results}

In this section, we present the numerical results of our model. We have performed a comprehensive scan over the parameter space defined in Sections~\ref{sec:scotogenic_model} and~\ref{Sec:mass-generation}, including neutrino Yukawa couplings, pseudo-Dirac masses, and charged scalar masses. For each parameter point, we impose consistency with the global fit to neutrino oscillation data from Ref.~\cite{Esteban:2024eli} at the $3\sigma$ level, as well as the stability and unitarity conditions derived in Appendix~\ref{sec:stability-unitarity}. Throughout this section, we present results for both normal (NO) and inverted (IO) neutrino mass orderings and examine their implications for charged lepton flavor violation and muonium-related observables.

\begin{figure}[htbp]
    \includegraphics[width=\textwidth]{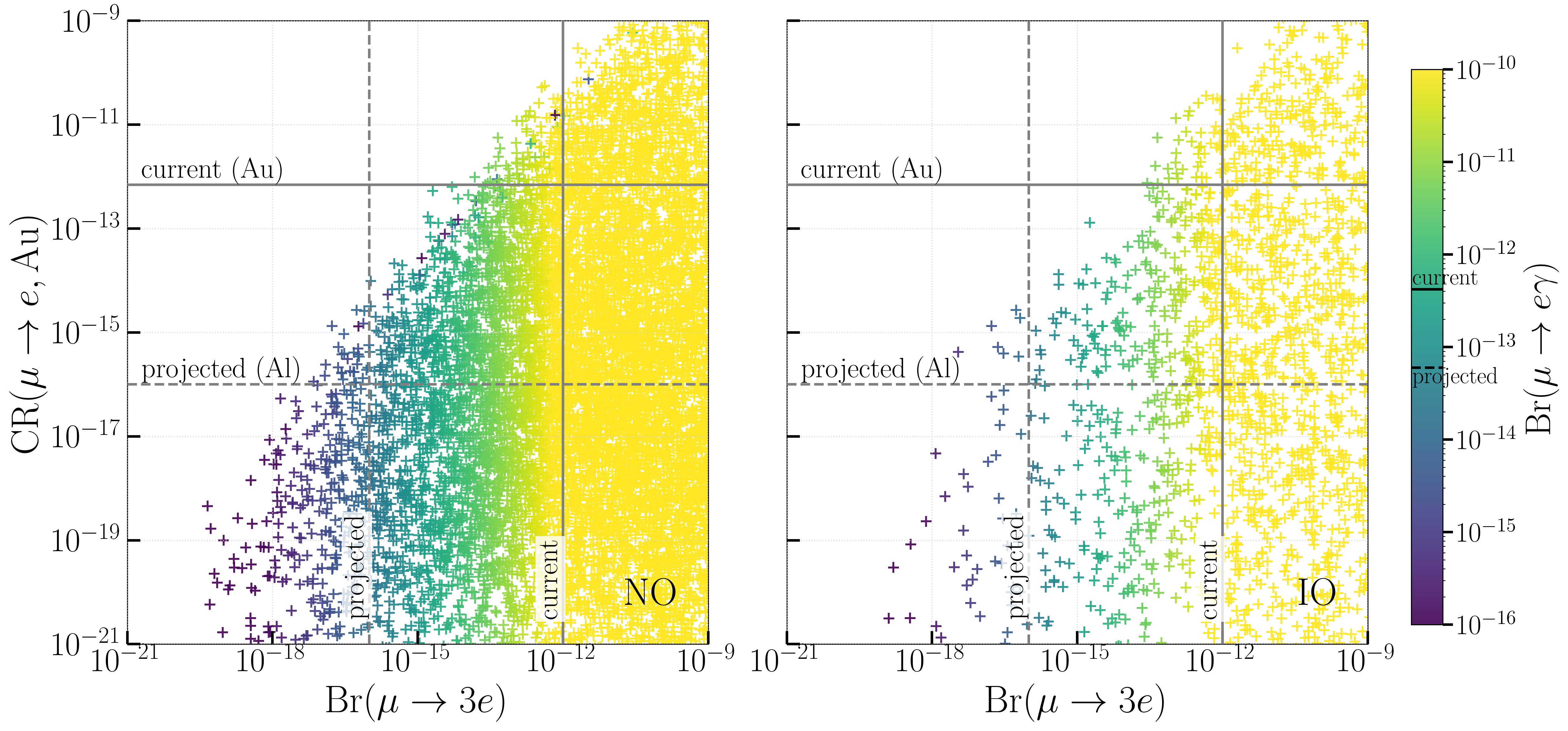}
    \captionsetup{justification=raggedright,singlelinecheck=false}
    \caption{Correlation among cLFV observables. Points allowed by the model and consistent with neutrino oscillation data assuming normal (inverted) ordering are shown in the left (right) panel. Current bounds are indicated by solid lines, while future projections are shown as dashed lines.}
    \label{fig:flavour_observables}
\end{figure}
We start with the most constraining cLFV observables in the muon sector. Figure \ref{fig:flavour_observables} shows the correlation between the $\mu\to e$ conversion rate in gold and the $\mu\to eee$ branching ratio for normal (left-panel) and inverted (right-panel) neutrino mass orderings. The color scale represents the corresponding $\mu\to e\gamma$ branching ratio. Our findings reveal a strong preference for the normal mass ordering over the inverted hierarchy, as it accommodates a vastly greater number of phenomenologically viable parameter points. 
There is also a clear trend where larger $\mu\to eee$ branching ratios imply larger $Br(\mu\to e\gamma)$. Out of the two contributions to cLFV, arising from the scalar and neutrino sectors, we have verified that the charged scalar mediators provide the dominant contribution due to the small mixings ($U_{\alpha4}$, $U_{\alpha5}$) required by neutrino data. Finally, it is also important to note that while the neutrino contributions to $\mu-e$ conversion and $\mu \to 3e$ are strongly correlated—since the same type of form factors contribute to both processes—this is no longer the case for the scalar contribution, which breaks the correlation.

\begin{center}
\begin{figure*}[htbp]
    \centering
    \includegraphics[width=1\textwidth]{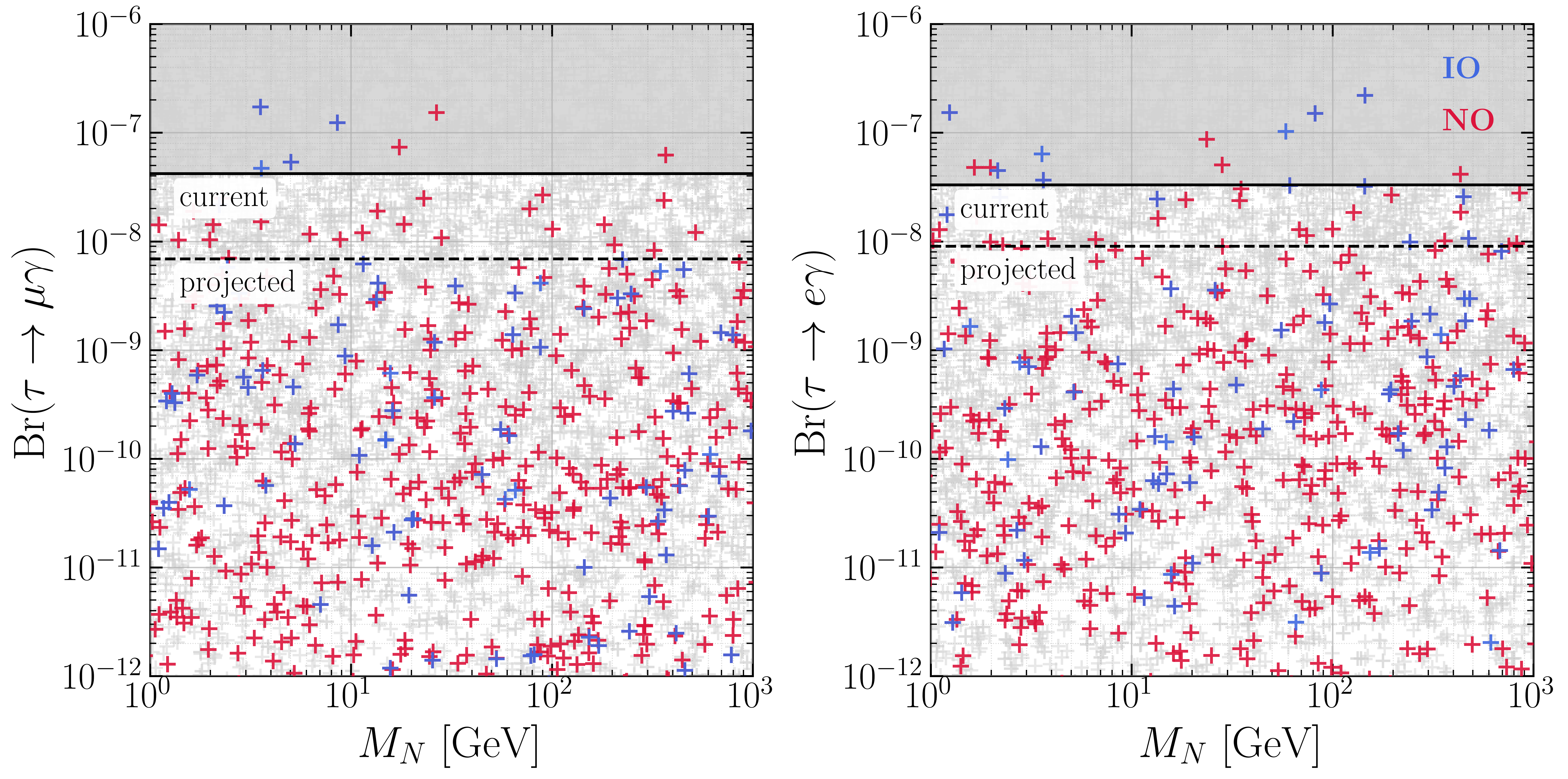}

   \caption{Most constraining cLFV $\tau$ decays are shown. In the left panel, we display $\mathrm{Br}(\tau \to \mu\gamma)$ as a function of $M_N$, and in the right panel, the equivalent plot for $\mathrm{Br}(\tau \to \mu\gamma)$. Red and blue crosses correspond to predictions consistent with normal (NO) and inverted (IO) neutrino mass orderings, respectively. Grey crosses indicate points excluded by other experimental constraints. The solid black lines denote the current experimental limits, while future projections are shown as dashed lines.}
    \label{fig:tau_decay}
\end{figure*}
\end{center}

\begin{figure}[htbp]
    \includegraphics[width=1\textwidth]{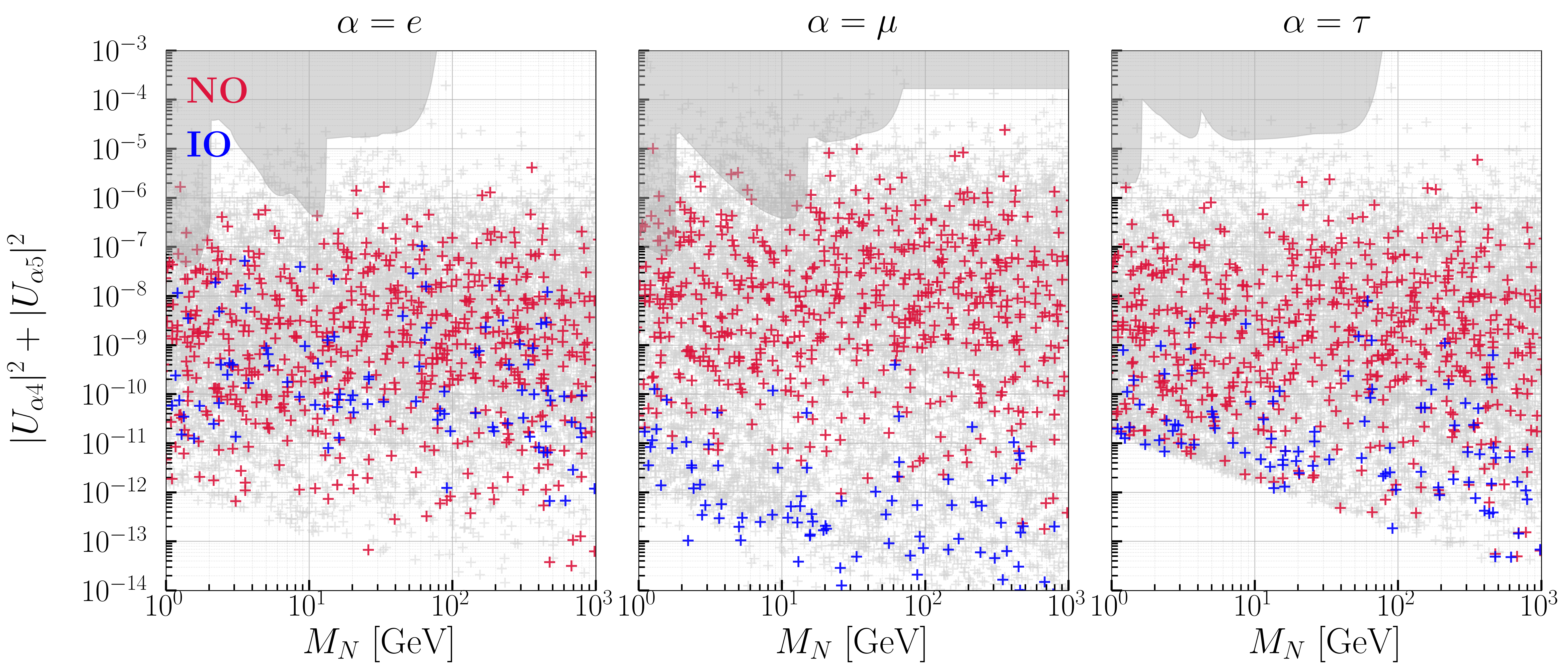}
    \captionsetup{justification=raggedright,singlelinecheck=false}
    \caption{Points consistent with neutrino oscillation data and allowed by the model. Red (blue) points correspond to normal (inverted) mass ordering. The gray shaded regions denote current experimental constraints\cite{Fernandez-Martinez:2023phj}, while gray points are excluded by existing cLFV constraints.}
    \label{fig:mixing_vs_mass}
\end{figure}

After analyzing the cLFV observables in the muon sector, it is natural to ask whether other cLFV processes, such as $\tau$ decays, can also provide relevant constraints. To address this question, Figure~\ref{fig:tau_decay} shows the most constraining branching ratios for $\tau$-sector cLFV, namely $\mathrm{Br}(\tau \to e\gamma)$ and $\mathrm{Br}(\tau \to \mu\gamma)$. 
The figure displays the points allowed by neutrino data, as well as the stability and unitarity conditions, for both normal (NO, red) and inverted (IO, blue) neutrino mass orderings. Points shown in grey correspond to regions already excluded by cLFV constraints from the muon sector. We observe that, after applying these constraints, the surviving points lie close to the current experimental bounds from $\tau$ decays, with a few additional points further excluded. Therefore, while $\tau$ decays are not as restrictive as those in the muon sector, they still provide complementary sensitivity and are not completely negligible.
 While scalar loops dominate the cLFV observables, the heavy neutrinos themselves are subject to direct search constraints. Figure~\ref{fig:mixing_vs_mass} shows the active–sterile mixing, quantified by $|U_{\alpha 4}|^2 + |U_{\alpha 5}|^2$ (with $\alpha = e, \mu, \tau$), as a function of the heavy neutrino mass $M_N$. All points displayed are consistent with neutrino oscillation data, as well as stability and unitarity conditions. Points excluded by cLFV observables are shown in grey, while the grey shaded regions indicate the exclusion limits from current direct search experiments. It is worth emphasizing that, although many points are ruled out by cLFV bounds, those processes are dominated by the scalar contribution. Therefore, when projected onto the plane of neutrino mixings, the allowed regions for large active–sterile mixing remain essentially unaffected.

\begin{figure}[htbp]
    \includegraphics[width=1\textwidth]{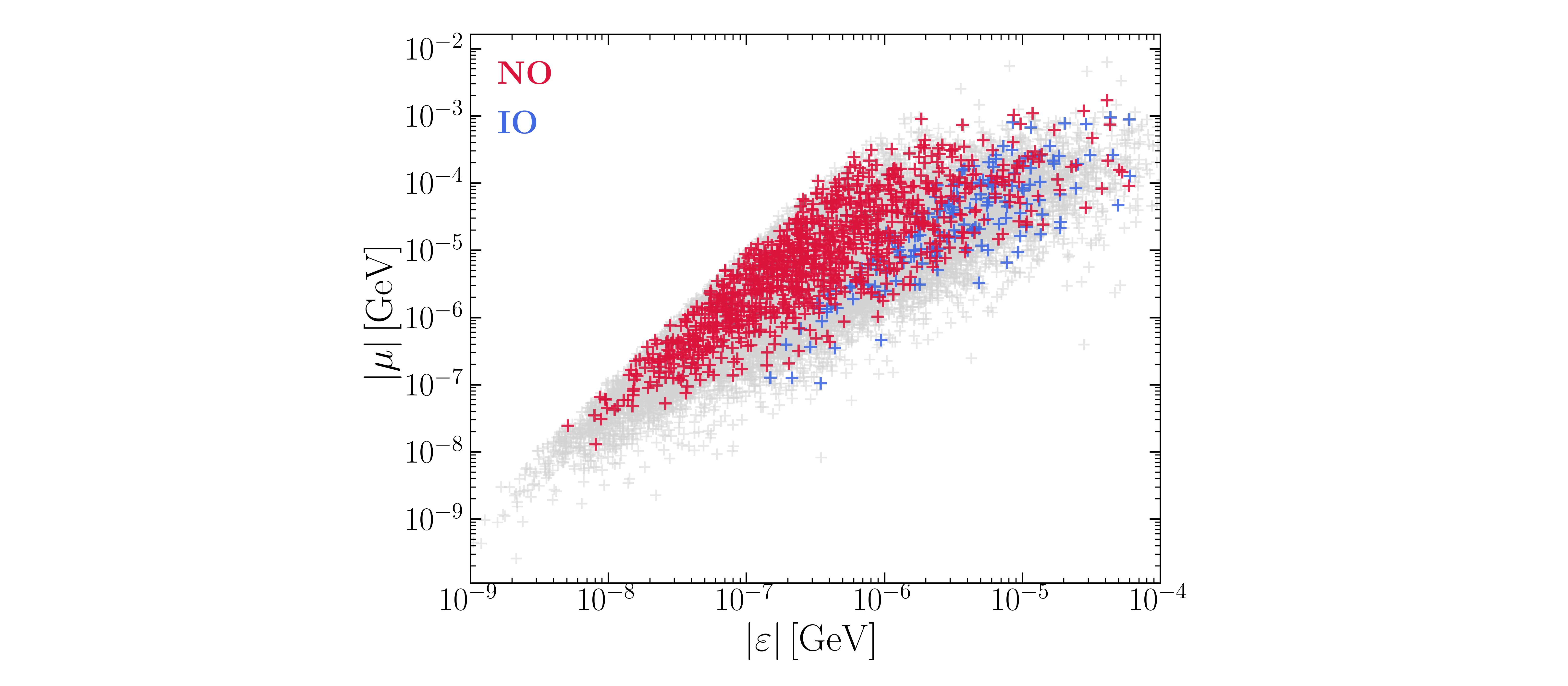}    \captionsetup{justification=raggedright,singlelinecheck=false}
    \caption{Correlation between the inverse seesaw parameter ($\mu$) and the linear seesaw term ($\varepsilon$). Points consistent with neutrino oscillation data and allowed by the model are shown. Red (blue) points correspond to normal (inverted) mass ordering, while gray points are excluded by current cLFV constraints.}
    \label{fig:epsilon_vs_mu}
\end{figure}

To further understand the neutrino mass generation mechanism in the model and the interplay between linear and inverse seesaw mechanisms, we show in Figure~\ref{fig:epsilon_vs_mu} the correlation between the inverse seesaw mass parameter $\mu$ and the linear seesaw parameter $\varepsilon$, where all points shown are consistent with neutrino oscillation experimental data. The red and blue points are compatible with charged lepton flavor violating constraints and correspond to the scenarios of normal and inverted neutrino mass orderings, respectively. We have found that the lepton number violating mass parameter $\mu$ is in general larger than the linear seesaw parameter $\varepsilon$. This suppression can be understood directly from the orthogonality conditions in Eq.~\eqref{eq:orthogonal_relations} which provides partial cancellation for $\varepsilon_i$ ($i=1,2.3$) explaining its suppression. This slight preference for the inverse seesaw is consistent with the underlying mechanisms, as the inverse seesaw contribution scales quadratically with the $\frac{m_D}{M}$ ratio, while the linear seesaw scales linearly. This analysis shows that the atmospheric mass squared splitting is generated from the inverse seesaw mechanism, whereas the solar mass squared difference arises from the linear seesaw.

Finally, we comment on the model predictions for Muonium observables. The cLFV decay $\text{Mu}\to e^+ e^-$ and the Mu–$\overline{\text{Mu}}$ conversion process are both controlled by the same flavour-violating couplings that appear in the decays $\mu \to e\gamma$ and $\mu \to 3e$. These latter processes provide much stronger experimental limits. As a consequence, after applying all cLFV constraints, we obtain
\begin{equation}
\text{BR}(\text{Mu} \to e^+ e^-) \sim 10^{-18} \text{--} 10^{-20},
\end{equation}
far below the current bounds discussed in Sec.~\ref{sec:muonium}, and still many orders of magnitude below future sensitivities. Likewise, Mu–$\overline{\text{Mu}}$ conversion, driven by loop-induced box diagrams with heavy neutrinos and dark scalars, leads to
\begin{equation}
|G_{M\overline{M}}| \lesssim 10^{-12} G_F ,
\end{equation}
corresponding to $P(\text{Mu}\to\overline{\text{Mu}}) < 10^{-20}$ and, like the case of Muonium decay, far below future sensitivity. We therefore conclude that Muonium-based searches are not expected to provide competitive probes of this model, whereas cLFV observables in the muon sector remain the most sensitive and promising experimental tests.

\section{Conclusions}

\label{Sec:conclusions}

We have proposed an economical and testable scotogenic model where the smallness of active neutrino masses is explained by a synergy of the linear and inverse seesaw mechanisms, which we name the LISS mechanism. The model extends the SM with a $U(1)'$ gauge symmetry and the discrete $\mathbb{Z}_3 \otimes \mathbb{Z}_4$ symmetries. The key lepton number violating parameters of the LSS ($\varepsilon$) and ISS ($\mu$) are not put in by hand but are dynamically generated at the two-loop level, naturally explaining their smallness. The preserved $\widetilde{\mathbb{Z}}_{2}\otimes\mathbb{Z}_{3}$ symmetries (where $\widetilde{\mathbb{Z}}_{2}$ arises from $\mathbb{Z}_4$ symmetry breaking) 
that enforce the radiative nature of the seesaw mechanism also guarantee the stability of both scalar and fermionic dark matter candidates, providing a unified framework for two of the most significant shortcomings of the Standard Model. Thanks to the symmetries, our model naturally predicts a fermionic dark matter candidate, i.e., $\Psi_R$, which is the lightest non trivial $Z_3$ particle, whose mass arises from a one loop level radiative seesaw mechanism induced by the virtual exchange of the Dark Dirac neutral leptons $\Omega_L$, $\Omega_R$ and the Dark scalar fields $\xi$, $\chi$. In contrast, the dark scalar masses arise at tree level.

Our phenomenological analysis shows that the model is consistent with all current experimental constraints, including neutrino oscillation data, direct searches for heavy neutrinos and $Z'$ bosons, as well as stringent bounds on cLFV processes. An interesting outcome of the analysis is that the inverse seesaw contribution dominates over the linear seesaw one, then implying that atmospheric neutrino mass squared splitting arises from the inverse seesaw mechanism whereas the solar mass squared difference is generated from the linear seesaw. This allows for an explanation of the hierarchy between the atmospheric and solar neutrino mass squared splittings, in addition to the smallness of active neutrino masses, feature not presented in many low-scale seesaw models. Moreover, our parameter-space scan shows a slight preference for the normal ordering, which exhibits a larger allowed region compared to the inverted ordering.

The model predicts cLFV rates—particularly for $\mu \to e\gamma$, $\mu \to eee$, and $\mu$–$e$ conversion—that lie within the reach of next-generation experiments such as MEG II, Mu3e, COMET, and Mu2e. We find that these processes are predominantly mediated by scalar loops involving $\eta$, since the neutrino contribution is strongly suppressed by the mixing required to satisfy neutrino data. Furthermore, we have explored distinctive predictions for Muonium–antimuonium oscillations and the decay $\text{Mu} \to e^+e^-$. However, the rates for these $|\Delta L|=2$ processes are well below current experimental limits and foreseeable future dedicated searches.

In conclusion, this scotogenic LISS model provides a compelling and verifiable explanation for the origin of neutrino masses, connecting it to dark matter and offering rich phenomenology at both the intensity and energy frontiers. Future work will involve a detailed study of the dark matter candidates and the potential for leptogenesis.


\section*{Acknowledgments}

 AECH is supported by ANID-Chile FONDECYT 1210378, 1241855, ANID PIA/APOYO AFB230003 and ANID – Millennium Science Initiative Program $ICN2019\_044$. This project has received funding from the European Union’s Horizon Europe research and innovation programme under the Marie Skłodowska-Curie Staff Exchange grant agreement No 101086085 – ASYMMETRY and support
from the IN2P3 (CNRS) Master Project, “Hunting for Heavy Neutral Leptons” (12-PH-0100).

\appendix

\section{Anomaly Cancellation}\label{sec:anom.canc}
By means of the conditions listed below, we have checked the cancellation of both gauge and gravitational anomalies:
\begin{align}
    A_{\left[ SU\left( 3\right) _{C}\right] ^{2}U\left( 1\right) '} 
    &= 2\sum_{i=1}^{3}\left( Q'\right) _{q_{iL}}
    - \sum_{i=1}^{3}\Big[ \left(Q'\right) _{u_{iR}}+\left( Q'\right) _{d_{iR}}\Big], \label{anom1}\\[1mm]
    A_{\left[ SU\left( 2\right) _{L}\right] ^{2}U\left( 1\right) '} 
    &= 2\sum_{i=1}^{3}\left( Q'\right) _{l_{iL}} + 6\sum_{i=1}^{3}\left( Q'\right) _{q_{iL}}, \label{anom2}\\[1mm]
    A_{\left[ U\left( 1\right) _{Y}\right] ^{2}U\left( 1\right) '} 
    &= 2\sum_{i=1}^{3}\Big[ Y_{l_{iL}}^{2}\left( Q'\right)_{l_{iL}}+3\,Y_{q_{iL}}^{2}\left( Q'\right) _{q_{iL}}\Big] \notag \\
    &\quad - \sum_{i=1}^{3}\Big[ Y_{l_{iR}}^{2}\left( Q'\right)_{l_{iR}}+3\,Y_{u_{iR}}^{2}\left( Q'\right) _{u_{iR}}+3\,Y_{d_{iR}}^{2}\left( Q'\right) _{d_{iR}}\Big], \label{anom3}\\[1mm]
    A_{\left[ U\left( 1\right) '\right] ^{2}U\left( 1\right) _{Y}} 
    &= 2\sum_{i=1}^{3}\Big[ Y_{l_{iL}}\left( Q'\right)_{l_{iL}}^{2}+3\,Y_{q_{iL}}\left( Q'\right) _{q_{iL}}^{2}\Big] \notag \\
    &\quad - \sum_{i=1}^{3}\Big[ Y_{l_{iR}}\left( Q'\right)_{l_{iR}}^{2}+3\,Y_{u_{iR}}\left( Q'\right) _{u_{iR}}^{2}+3\,Y_{d_{iR}}\left( Q'\right) _{d_{iR}}^{2}\Big], \label{anom4}\\[1mm]
    A_{\left[ U\left( 1\right) '\right] ^{3}} 
    &= \sum_{i=1}^{3}\Big[2\,(Q')_{l_{iL}}^{3}+6\,(Q')_{q_{iL}}^{3}-3\,(Q')_{u_{iR}}^{3}-3\,(Q')_{d_{iR}}^{3}\Big] \notag \\
    &\quad - \Biggl\{\sum_{i=1}^{3}(Q')_{l_{iR}}^{3} + (Q')_{\nu_{R}}^{3}\Biggr\} \notag \\
    &\quad + \Biggl\{ (Q')_{\Psi_L}^{3}+(Q')_{\Omega_L}^{3}\Biggr\}
    - \Biggl\{ (Q')_{N_R}^{3}+(Q')_{\Psi_{R}}^{3}+(Q')_{\tilde{\Psi}_{R}}^{3}+(Q')_{\Omega_R}^{3}\Biggr\}, \label{anom5}\\[1mm]
    A_{\left[ \text{Gravity}\right] ^{2}U\left( 1\right) '} 
    &= \sum_{i=1}^{3}\Big[2\,(Q')_{l_{iL}}+6\,(Q')_{q_{iL}}-3\,(Q')_{u_{iR}}-3\,(Q')_{d_{iR}}\Big] \notag \\
    &\quad - \Biggl\{\sum_{i=1}^{3}(Q')_{l_{iR}} + (Q')_{\nu_{R}}\Biggr\} \notag \\
    &\quad + \Biggl\{ (Q')_{\Psi_L}+(Q')_{\Omega_L}\Biggr\}
    - \Biggl\{ (Q')_{N_R}+(Q')_{\Psi_{R}}+(Q')_{\tilde{\Psi}_{R}}+(Q')_{\Omega_R}\Biggr\}. \label{anom6}
\end{align}
The hypercharge $Y$ and $U(1)'$ charges $Q'$ are given in Table~\ref{model}. 
In these expressions, $A$ denotes the anomaly coefficient obtained from the triangular diagrams with the gauge bosons corresponding to the groups in the subindices.

\section{Stability and unitarity conditions}
\label{sec:stability-unitarity}

In the regime of very large field values the behavior of the scalar potential is dominated by its quartic interactions. In order to study the stability of the potential, we first extract the quartic part, $V_4$, from the full scalar potential. To make our notation more compact, we introduce the following bilinear combinations:
\begin{equation}
\begin{aligned}
a &\equiv \phi^\dagger\phi,\quad b \equiv \sigma^\ast\sigma,\quad c \equiv \rho^2,\\[1mm]
d &\equiv \eta^\dagger\eta,\quad e \equiv \varphi^\ast\varphi,\quad f \equiv \xi^\ast\xi,\quad g \equiv \chi^\ast\chi\,.
\end{aligned}
\end{equation}
In addition, note that the term 
\begin{equation}
\lambda'_{\phi\eta}\left(\phi\eta\right)\left(\eta^\dagger\phi^\dagger\right)
\end{equation}
can be written as $\lambda'_{\phi\eta}\,|\phi^\dagger\eta|^2$. Furthermore, for the mixed term involving four fields we define
\begin{equation}
h \equiv \xi\sigma^{\ast}\chi\rho + \text{h.c.}\,.
\end{equation}
In terms of these bilinears the quartic part of the potential is given by
\allowdisplaybreaks
\begin{equation}
\begin{aligned}
V_4 &= \left( \sqrt{\lambda_{\phi}}\, a - \sqrt{\lambda_{\sigma}}\, b \right)^2 + \left( \sqrt{\lambda_{\phi}}\, a - \sqrt{\lambda_{\rho}}\, c \right)^2 + \left( \sqrt{\lambda_{\phi}}\, a - \sqrt{\lambda_{\eta}}\, d \right)^2 \\
&\quad + \left( \sqrt{\lambda_{\phi}}\, a - \sqrt{\lambda_{\varphi}}\, e \right)^2 + \left( \sqrt{\lambda_{\phi}}\, a - \sqrt{\lambda_{\xi}}\, f \right)^2 + \left( \sqrt{\lambda_{\phi}}\, a - \sqrt{\lambda_{\chi}}\, g \right)^2 \\
&\quad + \left( \sqrt{\lambda_{\sigma}}\, b - \sqrt{\lambda_{\rho}}\, c \right)^2 + \left( \sqrt{\lambda_{\sigma}}\, b - \sqrt{\lambda_{\eta}}\, d \right)^2 + \left( \sqrt{\lambda_{\sigma}}\, b - \sqrt{\lambda_{\varphi}}\, e \right)^2 \\
&\quad + \left( \sqrt{\lambda_{\sigma}}\, b - \sqrt{\lambda_{\xi}}\, f \right)^2 + \left( \sqrt{\lambda_{\sigma}}\, b - \sqrt{\lambda_{\chi}}\, g \right)^2 \\
&\quad + \left( \sqrt{\lambda_{\rho}}\, c - \sqrt{\lambda_{\eta}}\, d \right)^2 + \left( \sqrt{\lambda_{\rho}}\, c - \sqrt{\lambda_{\varphi}}\, e \right)^2 + \left( \sqrt{\lambda_{\rho}}\, c - \sqrt{\lambda_{\xi}}\, f \right)^2 \\
&\quad + \left( \sqrt{\lambda_{\rho}}\, c - \sqrt{\lambda_{\chi}}\, g \right)^2 \\
&\quad + \left( \sqrt{\lambda_{\eta}}\, d - \sqrt{\lambda_{\varphi}}\, e \right)^2 + \left( \sqrt{\lambda_{\eta}}\, d - \sqrt{\lambda_{\xi}}\, f \right)^2 + \left( \sqrt{\lambda_{\eta}}\, d - \sqrt{\lambda_{\chi}}\, g \right)^2 \\
&\quad + \left( \sqrt{\lambda_{\varphi}}\, e - \sqrt{\lambda_{\xi}}\, f \right)^2 + \left( \sqrt{\lambda_{\varphi}}\, e - \sqrt{\lambda_{\chi}}\, g \right)^2 + \left( \sqrt{\lambda_{\xi}}\, f - \sqrt{\lambda_{\chi}}\, g \right)^2  \\
&\quad -5\Bigl(\lambda_{\phi}\, a^2+\lambda_{\sigma}\, b^2+\lambda_{\rho}\, c^2+\lambda_{\eta}\, d^2+\lambda_{\varphi}\, e^2+\lambda_{\xi}\, f^2+\lambda_{\chi}\, g^2\Bigr) \\
&\quad + \lambda'_{\phi\eta}\, |\phi^\dagger\eta|^2 + \lambda_{\xi\sigma\chi\rho}\, h \\
&\quad + \Bigl( \lambda_{\phi\sigma} - 2\sqrt{\lambda_{\phi}\lambda_{\sigma}} \Bigr) a b + \Bigl( \lambda_{\phi\rho} - 2\sqrt{\lambda_{\phi}\lambda_{\rho}} \Bigr) a c \\
&\quad + \Bigl( \lambda_{\phi\eta} - 2\sqrt{\lambda_{\phi}\lambda_{\eta}} \Bigr) a d + \Bigl( \lambda_{\phi\varphi} - 2\sqrt{\lambda_{\phi}\lambda_{\varphi}} \Bigr) a e \\
&\quad + \Bigl( \lambda_{\phi\xi} - 2\sqrt{\lambda_{\phi}\lambda_{\xi}} \Bigr) a f + \Bigl( \lambda_{\phi\chi} - 2\sqrt{\lambda_{\phi}\lambda_{\chi}} \Bigr) a g \\
&\quad + \Bigl( \lambda_{\rho\sigma} - 2\sqrt{\lambda_{\sigma}\lambda_{\rho}} \Bigr) b c + \Bigl( \lambda_{\sigma\eta} - 2\sqrt{\lambda_{\sigma}\lambda_{\eta}} \Bigr) b d \\
&\quad + \Bigl( \lambda_{\sigma\varphi} - 2\sqrt{\lambda_{\sigma}\lambda_{\varphi}} \Bigr) b e + \Bigl( \lambda_{\sigma\xi} - 2\sqrt{\lambda_{\sigma}\lambda_{\xi}} \Bigr) b f \\
&\quad + \Bigl( \lambda_{\sigma\chi} - 2\sqrt{\lambda_{\sigma}\lambda_{\chi}} \Bigr) b g \\
&\quad + \Bigl( \lambda_{\rho\eta} - 2\sqrt{\lambda_{\rho}\lambda_{\eta}} \Bigr) c d + \Bigl( \lambda_{\rho\varphi} - 2\sqrt{\lambda_{\rho}\lambda_{\varphi}} \Bigr) c e \\
&\quad + \Bigl( \lambda_{\rho\xi} - 2\sqrt{\lambda_{\rho}\lambda_{\xi}} \Bigr) c f + \Bigl( \lambda_{\rho\chi} - 2\sqrt{\lambda_{\rho}\lambda_{\chi}} \Bigr) c g \\
&\quad + \Bigl( \lambda_{\eta\varphi} - 2\sqrt{\lambda_{\eta}\lambda_{\varphi}} \Bigr) d e + \Bigl( \lambda_{\eta\xi} - 2\sqrt{\lambda_{\eta}\lambda_{\xi}} \Bigr) d f \\
&\quad + \Bigl( \lambda_{\eta\chi} - 2\sqrt{\lambda_{\eta}\lambda_{\chi}} \Bigr) d g \\
&\quad + \Bigl( \lambda_{\varphi\xi} - 2\sqrt{\lambda_{\varphi}\lambda_{\xi}} \Bigr) e f + \Bigl( \lambda_{\varphi\chi} - 2\sqrt{\lambda_{\varphi}\lambda_{\chi}} \Bigr) e g \\
&\quad + \Bigl( \lambda_{\xi\chi} - 2\sqrt{\lambda_{\xi}\lambda_{\chi}} \Bigr) f g\,.
\end{aligned}
\end{equation}

Following the procedure outlined in Refs.~\cite{Maniatis:2006fs, Bhattacharyya:2015nca} for stability analysis, we conclude that our scalar potential is stable if the following conditions are met:
\begin{equation}
\lambda_{\phi},\,\lambda_{\sigma},\,\lambda_{\rho},\,\lambda_{\eta},\,\lambda_{\varphi},\,\lambda_{\xi},\,\lambda_{\chi},\,\lambda_{\xi\sigma\chi\rho},\,\lambda_{\phi\eta}' \geq 0,
\end{equation}
\begin{equation}
\begin{aligned}
\lambda_{\phi\sigma} + 2\sqrt{\lambda_{\phi}\lambda_{\sigma}} &\geq 0, \quad \lambda_{\phi\rho} + 2\sqrt{\lambda_{\phi}\lambda_{\rho}} \geq 0, \quad \lambda_{\phi\eta} + 2\sqrt{\lambda_{\phi}\lambda_{\eta}} \geq 0, \\
\lambda_{\phi\varphi} + 2\sqrt{\lambda_{\phi}\lambda_{\varphi}} &\geq 0, \quad \lambda_{\phi\xi} + 2\sqrt{\lambda_{\phi}\lambda_{\xi}} \geq 0, \quad \lambda_{\phi\chi} + 2\sqrt{\lambda_{\phi}\lambda_{\chi}} \geq 0, \\
\lambda_{\sigma\rho} + 2\sqrt{\lambda_{\sigma}\lambda_{\rho}} &\geq 0, \quad \lambda_{\sigma\eta} + 2\sqrt{\lambda_{\sigma}\lambda_{\eta}} \geq 0, \quad \lambda_{\sigma\varphi} + 2\sqrt{\lambda_{\sigma}\lambda_{\varphi}} \geq 0, \\
\lambda_{\sigma\xi} + 2\sqrt{\lambda_{\sigma}\lambda_{\xi}} &\geq 0, \quad \lambda_{\sigma\chi} + 2\sqrt{\lambda_{\sigma}\lambda_{\chi}} \geq 0, \quad  \lambda_{\rho\eta} + 2\sqrt{\lambda_{\rho}\lambda_{\eta}} \geq 0, \\
 \quad \lambda_{\rho\varphi} + 2\sqrt{\lambda_{\rho}\lambda_{\varphi}} &\geq 0, \quad\lambda_{\rho\xi} + 2\sqrt{\lambda_{\rho}\lambda_{\xi}} \geq 0,\quad \lambda_{\rho\chi} + 2\sqrt{\lambda_{\rho}\lambda_{\chi}} \geq 0.
\end{aligned}
\end{equation}

In addition to ensuring that the scalar potential is bounded from below, perturbative unitarity at high energies imposes further restrictions on the quartic couplings of the model. these unitary conditions are enforced to prevent scattering amplitudes involving physical scalar particles and longitudinal gauge bosons from increasing excessively with energy. A general amplitude $\mathcal{M}(\theta)$ may be expanded in partial waves as
\begin{equation}
\mathcal{M}(\theta) = 16\pi\sum_{\ell=0}^{\infty}(2\ell+1)a_{\ell}P_{\ell}(\cos\theta)\,,
\end{equation}
where $P_{\ell}(\cos\theta)$ are the Legendre polynomials and $a_{\ell}$ denotes the partial wave amplitude of order $\ell$. At energies where the scalar field amplitudes become large, the quartic interactions dominate the behavior of the potential. In this regime, one can study the two-body scattering processes involving the scalar fields, applying the equivalence theorem~\cite{Cornwall:1974km, Vayonakis:1976vz, Lee:1977eg, Gounaris:1986cr} to replace the longitudinal gauge boson modes by their corresponding Goldstone bosons.  Since the quartic interactions yield constant contributions at high energies, only the $s$-wave component, $a_0$, is significant. The unitarity condition is then enforced by requiring
\begin{equation}
|a_0| < 1\,,
\end{equation}
or, equivalently, that the eigenvalues of the relevant scattering matrices do not exceed $16\pi$ when the conventional normalization is adopted. We will follow a procedure analogous to that in Ref.~\cite{Bhattacharyya:2015nca,Abada:2021yot}, separating the $2 \rightarrow 2$ scattering processes into those involving electrically neutral pairs and those involving electrically charged pairs.

We will begin with the electrically neutral pairs. Due to the daunting number of scalar fields for clarity, we define the neutral scalar basis by dividing it into two distinct subbases:
\begin{equation}
     \mathcal{B}_A = \left(\phi^{-} \phi^{+},\eta^{-} \eta^{+},\frac{hh}{\sqrt2},\, \frac{\phi_Z\phi_Z}{\sqrt2},\, \frac{\tilde\sigma\,\tilde\sigma}{\sqrt2},\, \frac{\sigma_{Z'}\sigma_{Z'}}{\sqrt2},\, \frac{\tilde\rho\,\tilde\rho}{\sqrt2}\right)
\end{equation}
and
\begin{equation}
    \mathcal{B}_I =\left(\frac{\eta_R\eta_R}{\sqrt2},\, \frac{\eta_I\eta_I}{\sqrt2},\, \eta_R\eta_I,\, \frac{\varphi_R\varphi_R}{\sqrt2},\, \frac{\varphi_I\varphi_I}{\sqrt2},\, \varphi_R\varphi_I,\, \frac{\xi_R\xi_R}{\sqrt2},\, \frac{\xi_I\xi_I}{\sqrt2},\, \xi_R\xi_I,\, \frac{\chi_R\chi_R}{\sqrt2},\, \frac{\chi_I\chi_I}{\sqrt2},\, \chi_R\chi_I\right)\,.
\end{equation}
where their union forms the complete neutral scalar field basis:
\begin{equation} \mathcal{B}_N = \mathcal{B}_A \cup \mathcal{B}_I. \end{equation}
then the neutral scattering matrix, expressed in this basis, adopts the following block form:
\begin{equation}
\mathcal{S}_N =
\begin{pmatrix}
S_A & S_{AI}\\[1mm]
S_{AI}^T & S_I
\end{pmatrix}\,,
\end{equation}
where the explicit forms of the submatrices $S_A, S_{A I}$, and $S_I$ in terms of scalar couplings are provided respectively in equations (\ref{eq:SA}),(\ref{eq:SI}) and (\ref{eq:SAI}):
\begin{equation}
S_A = 
{\footnotesize
\left(\begin{array}{ccccccc}
4\lambda_\phi & \sqrt{2}(\lambda_{\phi\eta}+\lambda'_{\phi\eta}) & \sqrt{2}\lambda_\phi & \sqrt{2}\lambda_\phi & \frac{\lambda_{\phi\sigma}}{\sqrt{2}} & \frac{\lambda_{\phi\sigma}}{\sqrt{2}} & \frac{\lambda_{\phi\rho}}{\sqrt{2}} \\
\sqrt{2}(\lambda_{\phi\eta}+\lambda'_{\phi\eta}) & 4\lambda_\eta & \frac{\lambda_{\phi\eta}+\lambda'_{\phi\eta}}{\sqrt{2}} & \frac{\lambda_{\phi\eta}+\lambda'_{\phi\eta}}{\sqrt{2}} & \frac{\lambda_{\sigma\eta}}{\sqrt{2}} & \frac{\lambda_{\sigma\eta}}{\sqrt{2}} & \frac{\lambda_{\rho\eta}}{\sqrt{2}} \\
\sqrt{2}\lambda_\phi & \frac{\lambda_{\phi\eta}+\lambda'_{\phi\eta}}{\sqrt{2}} & 3\lambda_\phi & \lambda_\phi & \frac{\lambda_{\phi\sigma}}{2} & \frac{\lambda_{\phi\sigma}}{2} & \frac{\lambda_{\phi\rho}}{2} \\
\sqrt{2}\lambda_\phi & \frac{\lambda_{\phi\eta}+\lambda'_{\phi\eta}}{\sqrt{2}} & \lambda_\phi & 3\lambda_\phi & \frac{\lambda_{\phi\sigma}}{2} & \frac{\lambda_{\phi\sigma}}{2} & \frac{\lambda_{\phi\rho}}{2} \\
\frac{\lambda_{\phi\sigma}}{\sqrt{2}} & \frac{\lambda_{\sigma\eta}}{\sqrt{2}} & \frac{\lambda_{\phi\sigma}}{2} & \frac{\lambda_{\phi\sigma}}{2} & 3\lambda_\sigma & \lambda_\sigma & \lambda_{\rho\sigma} \\
\frac{\lambda_{\phi\sigma}}{\sqrt{2}} & \frac{\lambda_{\sigma\eta}}{\sqrt{2}} & \frac{\lambda_{\phi\sigma}}{2} & \frac{\lambda_{\phi\sigma}}{2} & \lambda_\sigma & 3\lambda_\sigma & \lambda_{\rho\sigma} \\
\frac{\lambda_{\phi\rho}}{\sqrt{2}} & \frac{\lambda_{\rho\eta}}{\sqrt{2}} & \frac{\lambda_{\phi\rho}}{2} & \frac{\lambda_{\phi\rho}}{2} & \lambda_{\rho\sigma} & \lambda_{\rho\sigma} & 4\lambda_\rho
\end{array}\right)
}.
\label{eq:SA}
\end{equation}

\begin{equation}
S_I = 
{\footnotesize
\left(\begin{array}{cccccccccccc}
4\lambda_\eta & 0 & \sqrt{2}\lambda_\eta & \frac{\lambda_{\eta\varphi}}{\sqrt{2}} & \frac{\lambda_{\eta\varphi}}{\sqrt{2}} & \frac{\lambda_{\eta\varphi}}{2} & \frac{\lambda_{\eta\xi}}{\sqrt{2}} & \frac{\lambda_{\eta\xi}}{\sqrt{2}} & \frac{\lambda_{\eta\xi}}{2} & \frac{\lambda_{\eta\chi}}{\sqrt{2}} & \frac{\lambda_{\eta\chi}}{\sqrt{2}} & \frac{\lambda_{\eta\chi}}{2} \\ 
0 & 4\lambda_\eta & \sqrt{2}\lambda_\eta & \frac{\lambda_{\eta\varphi}}{\sqrt{2}} & \frac{\lambda_{\eta\varphi}}{\sqrt{2}} & \frac{\lambda_{\eta\varphi}}{2} & \frac{\lambda_{\eta\xi}}{\sqrt{2}} & \frac{\lambda_{\eta\xi}}{\sqrt{2}} & \frac{\lambda_{\eta\xi}}{2} & \frac{\lambda_{\eta\chi}}{\sqrt{2}} & \frac{\lambda_{\eta\chi}}{\sqrt{2}} & \frac{\lambda_{\eta\chi}}{2} \\ 
\sqrt{2}\lambda_\eta & \sqrt{2}\lambda_\eta & 2\lambda_\eta & \frac{\lambda_{\eta\varphi}}{2} & \frac{\lambda_{\eta\varphi}}{2} & 0 & \frac{\lambda_{\eta\xi}}{2} & \frac{\lambda_{\eta\xi}}{2} & 0 & \frac{\lambda_{\eta\chi}}{2} & \frac{\lambda_{\eta\chi}}{2} & 0 \\ 
\frac{\lambda_{\eta\varphi}}{\sqrt{2}} & \frac{\lambda_{\eta\varphi}}{\sqrt{2}} & \frac{\lambda_{\eta\varphi}}{2} & 3\lambda_\varphi & \lambda_\varphi & \sqrt{2}\lambda_\varphi & \frac{\lambda_{\varphi\xi}}{2} & \frac{\lambda_{\varphi\xi}}{2} & \frac{\lambda_{\varphi\xi}}{\sqrt{2}} & \frac{\lambda_{\varphi\chi}}{\sqrt{2}} & \frac{\lambda_{\varphi\chi}}{\sqrt{2}} & \frac{\lambda_{\varphi\chi}}{2} \\ 
\frac{\lambda_{\eta\varphi}}{\sqrt{2}} & \frac{\lambda_{\eta\varphi}}{\sqrt{2}} & \frac{\lambda_{\eta\varphi}}{2} & \lambda_\varphi & 3\lambda_\varphi & \sqrt{2}\lambda_\varphi & \frac{\lambda_{\varphi\xi}}{2} & \frac{\lambda_{\varphi\xi}}{2} & \frac{\lambda_{\varphi\xi}}{\sqrt{2}} & \frac{\lambda_{\varphi\chi}}{\sqrt{2}} & \frac{\lambda_{\varphi\chi}}{\sqrt{2}} & \frac{\lambda_{\varphi\chi}}{2} \\ 
\frac{\lambda_{\eta\varphi}}{2} & \frac{\lambda_{\eta\varphi}}{2} & 0 & \sqrt{2}\lambda_\varphi & \sqrt{2}\lambda_\varphi & 2\lambda_\varphi & \frac{\lambda_{\varphi\xi}}{\sqrt{2}} & \frac{\lambda_{\varphi\xi}}{\sqrt{2}} & 0 & \frac{\lambda_{\varphi\chi}}{2} & \frac{\lambda_{\varphi\chi}}{2} & 0 \\ 
\frac{\lambda_{\eta\xi}}{\sqrt{2}} & \frac{\lambda_{\eta\xi}}{\sqrt{2}} & \frac{\lambda_{\eta\xi}}{2} & \frac{\lambda_{\varphi\xi}}{2} & \frac{\lambda_{\varphi\xi}}{2} & \frac{\lambda_{\varphi\xi}}{\sqrt{2}} & 3\lambda_\xi & \lambda_\xi & \sqrt{2}\lambda_\xi & \frac{\lambda_{\xi\chi}}{\sqrt{2}} & \frac{\lambda_{\xi\chi}}{\sqrt{2}} & \frac{\lambda_{\xi\chi}}{2} \\ 
\frac{\lambda_{\eta\xi}}{\sqrt{2}} & \frac{\lambda_{\eta\xi}}{\sqrt{2}} & \frac{\lambda_{\eta\xi}}{2} & \frac{\lambda_{\varphi\xi}}{2} & \frac{\lambda_{\varphi\xi}}{2} & \frac{\lambda_{\varphi\xi}}{\sqrt{2}} & \lambda_\xi & 3\lambda_\xi & \sqrt{2}\lambda_\xi & \frac{\lambda_{\xi\chi}}{\sqrt{2}} & \frac{\lambda_{\xi\chi}}{\sqrt{2}} & \frac{\lambda_{\xi\chi}}{2} \\ 
\frac{\lambda_{\eta\xi}}{2} & \frac{\lambda_{\eta\xi}}{2} & 0 & \frac{\lambda_{\varphi\xi}}{\sqrt{2}} & \frac{\lambda_{\varphi\xi}}{\sqrt{2}} & 0 & \sqrt{2}\lambda_\xi & \sqrt{2}\lambda_\xi & 2\lambda_\xi & \frac{\lambda_{\xi\chi}}{2} & \frac{\lambda_{\xi\chi}}{2} & 0 \\ 
\frac{\lambda_{\eta\chi}}{\sqrt{2}} & \frac{\lambda_{\eta\chi}}{\sqrt{2}} & \frac{\lambda_{\eta\chi}}{2} & \frac{\lambda_{\varphi\chi}}{\sqrt{2}} & \frac{\lambda_{\varphi\chi}}{\sqrt{2}} & \frac{\lambda_{\varphi\chi}}{2} & \frac{\lambda_{\xi\chi}}{\sqrt{2}} & \frac{\lambda_{\xi\chi}}{\sqrt{2}} & \frac{\lambda_{\xi\chi}}{2} & 3\lambda_\chi & \lambda_\chi & \sqrt{2}\lambda_\chi \\ 
\frac{\lambda_{\eta\chi}}{\sqrt{2}} & \frac{\lambda_{\eta\chi}}{\sqrt{2}} & \frac{\lambda_{\eta\chi}}{2} & \frac{\lambda_{\varphi\chi}}{\sqrt{2}} & \frac{\lambda_{\varphi\chi}}{\sqrt{2}} & \frac{\lambda_{\varphi\chi}}{2} & \frac{\lambda_{\xi\chi}}{\sqrt{2}} & \frac{\lambda_{\xi\chi}}{\sqrt{2}} & \frac{\lambda_{\xi\chi}}{2} & \lambda_\chi & 3\lambda_\chi & \sqrt{2}\lambda_\chi \\ 
\frac{\lambda_{\eta\chi}}{2} & \frac{\lambda_{\eta\chi}}{2} & 0 & \frac{\lambda_{\varphi\chi}}{2} & \frac{\lambda_{\varphi\chi}}{2} & 0 & \frac{\lambda_{\xi\chi}}{2} & \frac{\lambda_{\xi\chi}}{2} & 0 & \sqrt{2}\lambda_\chi & \sqrt{2}\lambda_\chi & 2\lambda_\chi
\end{array}\right)
}.
\label{eq:SI}
\end{equation}

\begin{equation}
S_{AI} = 
{\footnotesize
\left(\begin{array}{cccccccccccc}
\sqrt{2}(\lambda_{\phi\eta}+\lambda'_{\phi\eta}) & \sqrt{2}(\lambda_{\phi\eta}+\lambda'_{\phi\eta}) & (\lambda_{\phi\eta}+\lambda'_{\phi\eta}) & \dfrac{\lambda_{\phi\varphi}}{\sqrt{2}} & \dfrac{\lambda_{\phi\varphi}}{\sqrt{2}} & \dfrac{\lambda_{\phi\varphi}}{2} & \dfrac{\lambda_{\phi\xi}}{\sqrt{2}} & \dfrac{\lambda_{\phi\xi}}{\sqrt{2}} & \dfrac{\lambda_{\phi\xi}}{2} & \dfrac{\lambda_{\phi\chi}}{\sqrt{2}} & \dfrac{\lambda_{\phi\chi}}{\sqrt{2}} & \dfrac{\lambda_{\phi\chi}}{2} \\ 
\frac{1}{\sqrt{2}}(\lambda_{\phi\eta}+\lambda'_{\phi\eta}) & \frac{1}{\sqrt{2}}(\lambda_{\phi\eta}+\lambda'_{\phi\eta}) & \frac{1}{2}(\lambda_{\phi\eta}+\lambda'_{\phi\eta}) & \dfrac{\lambda_{\eta\varphi}}{\sqrt{2}} & \dfrac{\lambda_{\eta\varphi}}{\sqrt{2}} & \dfrac{\lambda_{\eta\varphi}}{2} & \dfrac{\lambda_{\eta\xi}}{\sqrt{2}} & \dfrac{\lambda_{\eta\xi}}{\sqrt{2}} & \dfrac{\lambda_{\eta\xi}}{2} & \dfrac{\lambda_{\eta\chi}}{\sqrt{2}} & \dfrac{\lambda_{\eta\chi}}{\sqrt{2}} & \dfrac{\lambda_{\eta\chi}}{2} \\ 
\sqrt{2}\lambda_{\phi} & \sqrt{2}\lambda_{\phi} & \lambda_{\phi} & \dfrac{\lambda_{\phi\varphi}}{\sqrt{2}} & \dfrac{\lambda_{\phi\varphi}}{\sqrt{2}} & \dfrac{\lambda_{\phi\varphi}}{2} & \dfrac{\lambda_{\phi\xi}}{\sqrt{2}} & \dfrac{\lambda_{\phi\xi}}{\sqrt{2}} & \dfrac{\lambda_{\phi\xi}}{2} & \dfrac{\lambda_{\phi\chi}}{\sqrt{2}} & \dfrac{\lambda_{\phi\chi}}{\sqrt{2}} & \dfrac{\lambda_{\phi\chi}}{2} \\ 
\sqrt{2}\lambda_{\phi} & \sqrt{2}\lambda_{\phi} & \lambda_{\phi} & \dfrac{\lambda_{\phi\varphi}}{\sqrt{2}} & \dfrac{\lambda_{\phi\varphi}}{\sqrt{2}} & \dfrac{\lambda_{\phi\varphi}}{2} & \dfrac{\lambda_{\phi\xi}}{\sqrt{2}} & \dfrac{\lambda_{\phi\xi}}{\sqrt{2}} & \dfrac{\lambda_{\phi\xi}}{2} & \dfrac{\lambda_{\phi\chi}}{\sqrt{2}} & \dfrac{\lambda_{\phi\chi}}{\sqrt{2}} & \dfrac{\lambda_{\phi\chi}}{2} \\ 
\dfrac{\lambda_{\sigma\eta}}{\sqrt{2}} & \dfrac{\lambda_{\sigma\eta}}{\sqrt{2}} & \dfrac{\lambda_{\sigma\eta}}{2} & \dfrac{\lambda_{\sigma\varphi}}{\sqrt{2}} & \dfrac{\lambda_{\sigma\varphi}}{\sqrt{2}} & \dfrac{\lambda_{\sigma\varphi}}{2} & \dfrac{\lambda_{\sigma\xi}}{\sqrt{2}} & \dfrac{\lambda_{\sigma\xi}}{\sqrt{2}} & \dfrac{\lambda_{\sigma\xi}}{2} & \dfrac{\lambda_{\sigma\chi}}{\sqrt{2}} & \dfrac{\lambda_{\sigma\chi}}{\sqrt{2}} & \dfrac{\lambda_{\sigma\chi}}{2} \\ 
\dfrac{\lambda_{\rho\eta}}{\sqrt{2}} & \dfrac{\lambda_{\rho\eta}}{\sqrt{2}} & \dfrac{\lambda_{\rho\eta}}{2} & \dfrac{\lambda_{\rho\varphi}}{\sqrt{2}} & \dfrac{\lambda_{\rho\varphi}}{\sqrt{2}} & \dfrac{\lambda_{\rho\varphi}}{2} & \dfrac{\lambda_{\rho\xi}}{\sqrt{2}} & \dfrac{\lambda_{\rho\xi}}{\sqrt{2}} & \dfrac{\lambda_{\rho\xi}}{2} & \dfrac{\lambda_{\rho\chi}}{\sqrt{2}} & \dfrac{\lambda_{\rho\chi}}{\sqrt{2}} & \dfrac{\lambda_{\rho\chi}}{2}
\end{array}\right)
}.
\label{eq:SAI}
\end{equation}

Similarly, we introduce the charged scalar basis organized in two subbases:
\begin{equation} \mathcal{B}_C^{\phi^{+}}=\left(\phi^{+} h, \phi^{+} \phi_Z, \phi^{+} \tilde{\sigma}, \phi^{+} \sigma_{Z^{\prime}}, \phi^{+} \tilde{\rho}, \phi^{+} \eta_R, \phi^{+} \eta_I, \phi^{+} \varphi_R, \phi^{+} \varphi_I, \phi^{+} \xi_R, \phi^{+} \xi_I, \phi^{+} \chi_R, \phi^{+} \chi_I\right), \end{equation}
and
\begin{equation} \mathcal{B}_C^{\eta^{-}} = \left(\eta^{-} h, \eta^{-} \phi_Z, \eta^{-} \tilde{\sigma}, \eta^{-} \sigma_{Z^{\prime}}, \eta^{-} \tilde{\rho}, \eta^{-} \eta_R, \eta^{-} \eta_I, \eta^{-} \varphi_R, \eta^{-} \varphi_I, \eta^{-} \xi_R, \eta^{-} \xi_I, \eta^{-}\chi_R, \eta^{-} \chi_I\right). \end{equation}

Their union forms the complete charged scalar field basis:

\begin{equation} \mathcal{B}_C = \mathcal{B}_C^{\phi^{+}} \cup \mathcal{B}_C^{\eta^{-}}. \end{equation}

In this charged scalar basis, the corresponding scattering matrix  $\mathcal{S}_C$  takes a block-diagonal form: 

\begin{equation}
\mathcal{S}_C =
\begin{pmatrix}
\mathcal{S}_C^{\phi^{+}} & 0\\[1mm]
0 & \mathcal{S}_C^{\eta^{-}}
\end{pmatrix}\,,
\end{equation}
where the submatrices $\mathcal{S}_C^{\phi^{+}},\mathcal{S}_C^{\eta^{-}}$ are given by equations (\ref{eq:S_phi}),(\ref{eq:S_eta}) respectively:

\begin{align}
\mathcal{S}_C^{\phi^{+}} &= \operatorname{diag}\Bigl(2\lambda_\phi,\, 2\lambda_\phi,\, \lambda_{\phi\sigma},\, \lambda_{\phi\sigma},\, \lambda_{\phi\rho},\, \lambda_{\phi\eta},\, \lambda_{\phi\eta},\, \lambda_{\phi\varphi},\, \lambda_{\phi\varphi},\, \lambda_{\phi\xi},\, \lambda_{\phi\xi},\, \lambda_{\phi\chi},\, \lambda_{\phi\chi}\Bigr) \label{eq:S_phi} \\[1ex]
\mathcal{S}_C^{\eta^{-}}  &= \operatorname{diag}\Bigl(2\lambda_\eta,\, 2\lambda_\eta,\, \lambda_{\eta\sigma},\, \lambda_{\eta\sigma},\, \lambda_{\eta\rho},\, \lambda_\eta,\, \lambda_\eta,\, \lambda_{\eta\varphi},\, \lambda_{\eta\varphi},\, \lambda_{\eta\xi},\, \lambda_{\eta\xi},\, \lambda_{\eta\chi},\, \lambda_{\eta\chi}\Bigr) \label{eq:S_eta}
\end{align}
The unitarity conditions require the moduli of the eigenvalues of  $\mathcal{S}_N$ and  $\mathcal{S}_C$ to be less than $16\pi$.

\section{Loop Functions and Form Factors}\label{app:C}

In this appendix, we collect the explicit expressions for the form factors and loop functions entering the charged lepton flavor violating observables discussed in Section~\ref{sec:clfv}. The form factors $F^{\mu e}_{\gamma}, F_Z^{\mu e}, G_\gamma^{\mu e},
F_{\rm Box}^{\mu e \alpha \alpha}$ are given by~\cite{Cvetic:2006yg}:
\begin{eqnarray}
\label{eq:FF}
F^{\mu e }_\gamma &=& \sum_{j=1}^{3 + n_S} {\bf U}_{ej}{\bf U}^*_{\mu
  j} F_\gamma(x_j)\,, \nonumber \label{Fmue}\\ 
G^{\mu e }_\gamma &=& \sum_{j=1}^{3 + n_S} {\bf U}_{ej}{\bf U}^*_{\mu
  j} G_\gamma(x_j)\,,  \nonumber\label{Ggammamue} \\ 
F^{\mu e }_Z &=& \sum_{j,k=1}^{3 + n_S} {\bf U}_{ej}{\bf U}^*_{\mu k}
\left(\delta_{jk} F_Z(x_j) + {\bf C}_{jk} G_Z(x_j,x_k) + {\bf
  C}^*_{jk} H_Z(x_j,x_k)   \right)\,, \nonumber \\ 
F^{\mu e uu}_{\rm Box}&=&\sum_{j=1}^{3 + n_S}\sum_{d_\alpha=d,s,b}
{\bf U}_{ej}{\bf U}^*_{\mu j} V_{u d_\alpha} V^*_{u d_\alpha} F_{\rm
  Box}(x_j,x_{d_\alpha})\,,   \nonumber \\
F^{\mu e dd}_{\rm Box}&=&  \sum_{j=1}^{3 + n_S}\sum_{u_\alpha=u,c,t}
{\bf U}_{ej}{\bf U}^*_{\mu j} V_{d  u_\alpha } V^*_{d u_\alpha} F_{\rm
  XBox}(x_j,x_{u_\alpha})\,,   \nonumber \\ 
F^{\mu eee}_{\rm Box}&=&  \sum_{j,k=1}^{3 + n_S}{\bf U}_{e j} {\bf
  U}^*_{\mu k}\left({\bf U}_{e j}{\bf U}^*_{ek}G_{\rm
  Box}(x_j,x_k)-2\,{\bf U}^*_{e j}{\bf U}_{e k}F_{\rm
  XBox}(x_j,x_k)\right)\,, \label{Fmueee}  
\end{eqnarray}
where $V_{q q^\prime}$ denotes the quark CKM matrix, 
${\bf C}_{jk} \equiv \sum_{\alpha = e,\mu,\tau} {\bf U}_{\alpha j}\,{\bf U}^{*}_{\alpha k}$, 
and $x_i = m_{\nu_i}^2/m_W^2$ is the dimensionless mass ratio.  In the limit of light masses ($x\ll 1$), 
the form factors assume the following asymptotic behavior:
\begin{align}
F^{\mu e }_\gamma &  \xrightarrow[x\ll 1]{}  \sum_{j=1}^{3 + n_S}{\bf
  U}_{e j}{\bf U}^*_{\mu j} \left[-x_{j}\right] \, &	  
 G^{\mu e }_\gamma &  \xrightarrow[x\ll 1]{} \sum_{j=1}^{3 + n_S} {\bf
   U}_{e j}{\bf U}^*_{\mu j} \left[ \frac{x_j}{4} \right] \, \nonumber
 \\ 
F^{\mu e }_Z & \xrightarrow[x\ll 1]{} \sum_{j=1}^{3 + n_S} {\bf U}_{e
  j}{\bf U}^*_{\mu j} \left[ x_{j} \left( \frac{-5}{2}- \ln x_{j}
  \right) \right] \, & 
F^{\mu eee}_{\rm  Box} & \xrightarrow[x\ll 1]{}  \sum_{j=1}^{3 + n_S}
{\bf U}_{e j} {\bf U}^*_{\mu j}\left[2~x_{j} \left(1+\ln x_{j}
  \right) \right] \, .   
\end{align}
Finally, the computation of the different amplitudes calls upon the
following loop
functions~\cite{Alonso:2012ji,Ilakovac:1994kj,Ma:1979px,Gronau:1984ct},
entering the form factors of Eq.~(\ref{Fmueee}):
\begin{align}
F_Z(x)&= -\frac{5x}{2(1-x)}-\frac{5x^2}{2(1-x)^2}\ln x \, , \nonumber
\\  
G_Z(x,y)&= -\frac{1}{2(x-y)}\left[	\frac{x^2(1-y)}{1-x}\ln x -
  \frac{y^2(1-x)}{1-y}\ln y	\right]\, ,  \nonumber \\ 
H_Z(x,y)&=  \frac{\sqrt{xy}}{4(x-y)}\left[	\frac{x^2-4x}{1-x}\ln
  x - \frac{y^2-4y}{1-y}\ln y	\right] \, , \nonumber \\ 
F_\gamma(x)&= 	\frac{x(7x^2-x-12)}{12(1-x)^3} -
\frac{x^2(x^2-10x+12)}{6(1-x)^4} \ln x	\, , \nonumber \\ 
G_\gamma(x)&=    -\frac{x(2x^2+5x-1)}{4(1-x)^3} -
\frac{3x^3}{2(1-x)^4} \ln x \, ,\label{Ggamma}  \nonumber \\	
F_{\rm Box}(x, y) &= \frac{1}{x - y} \bigg\{
\left(4+\frac{x  
y}{4}\right)\left[\frac{1}{1-x}+\frac{x^2}{(1-x)^2}\ln
x\right] - 2x 
y\left[\frac{1}{1-x}+\frac{x}{(1-x)^2}\ln
x\right] -(x\to y)\bigg\} \,,\nonumber \\
F_{\rm XBox}(x, y) &= \frac{-1}{x - y} \bigg\{
\left(1+\frac{x
y}{4}\right)\left[\frac{1}{1-x}+\frac{x^2}{(1-x)^2}\ln
x\right] - 2x
y\left[\frac{1}{1-x}+\frac{x}{(1-x)^2}\ln
x\right] -(x\to y)\bigg\} \,. 			
 \end{align}
In the limit of light masses ($x\ll 1$) and/or degenerate propagators
($x=y$), one has 
\begin{align}
F_Z(x) &  \xrightarrow[x\ll 1]{}    -\frac{5x}{2} \,,\nonumber\\
G_Z(x,x ) &= {} -\left[x (-1 + x - 2 \ln x)/(2 (x -1)) \right]\, , \,\,
G_Z(x,x)  \xrightarrow[x\ll 1]{}  -\frac{1}{2} x \ln x \,, \nonumber\\
H_Z(x,x ) & = {} - \left[ \sqrt{x^2} (4 - 5x + x^2 + (4 - 2x + x^2)\ln
  x)/(4(x - 1)^2) \right] \, , \nonumber\\ 
F_\gamma(x ) & \xrightarrow[x\ll 1]{}  -x \,, \nonumber\\
G_\gamma(x )  & \xrightarrow[x\ll 1]{}  \frac{x}{4}\, \nonumber\\
 F_{\rm Box}(x,x)	 &= \left[\left(-16 + 31x^2 - 16x^3 + x^4 +
   2x(-16 + 4x + 3x^2) \ln x \right)/\left(4(-1 + x)^3\right) \right]
  \, , \nonumber\\ 
F_{\rm XBox}(x,x )& = \left[ (-4 + 19 x^2 - 16 x^3 + x^4 + 2x (-4  4 x
  + 3x^2) \ln x)/(4(x - 1)^3) \right] \, .\label{limitval2} 
\end{align}

\section{One--Loop LFV Higgs Yukawa and Matching to Scalar Operators}
\label{app:LFV-Yukawa}

In this Appendix, we connect the fundamental parameters of our model to the effective quark–level scalar couplings $g_{\text{LS}}^{(q)}$ introduced in Eq.~\eqref{eq:nucleon_effective_couplings}, which enter in $\mu\to e$ conversion in nuclei.

LFV scalar operators are radiatively generated from the interactions in Eq.~\eqref{eq:Yukawa_lagrangian}, in particular
\begin{equation}
\mathcal{L} \supset (y_\Psi)_{i1}\,\overline{l_{iL}}\,\eta\,\Psi_{R} + \text{h.c.},
\end{equation}
together with the cubic scalar coupling from the scalar potential,
\begin{equation}
\mathcal{L} \supset -\,\kappa_{\eta\eta h}\,h\,\eta^{\dagger}\eta,
\qquad 
\kappa_{\eta\eta h} \equiv \frac{v_\phi}{\sqrt{2}}\bigl(\lambda_{\phi\eta}+\lambda'_{\phi\eta}\bigr),
\end{equation}
where $\phi^0 = (v_\phi + h)/\sqrt{2}$.

\begin{figure}[h!]
    \centering
\includegraphics[width=0.45\textwidth]{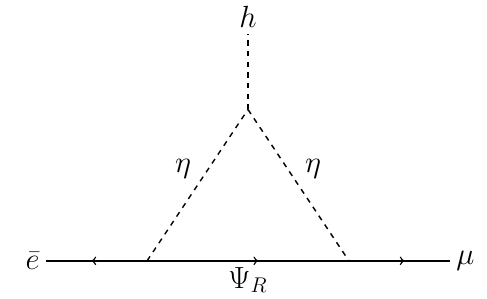}
    \caption{One--loop diagram inducing the $h\,\bar e\,\mu$ coupling through $(\eta,\Psi_R)$ exchange.}
    \label{fig:LFVtriangle}
\end{figure}

At one loop level, an effective LFV Higgs Yukawa coupling is induced by the triangle Feynman diagram 
with internal $(\eta,\,\Psi_R,\,\eta)$ propagators, as shown in Fig.~\ref{fig:LFVtriangle}. The required chirality flip occurs on the $\Psi_R$ internal line, then ensuring that the generated effective vertex is purely left–chiral.

The loop--induced interaction can be written as
\begin{equation}
\mathcal{L} \supset 
-\,\bar e\,\bigl(y^h_{\mu e,L}\,P_L + y^h_{\mu e,R}\,P_R\bigr)\,\mu\,h 
+ \text{h.c.},
\end{equation}
where the coefficients are given by
\begin{equation}
y^h_{\mu e,L} =
\frac{(y_\Psi)_{\mu1}(y_\Psi)_{e1}}{16\pi^2}\;
\kappa_{\eta\eta h}\;
m_\Psi\;
C_0(0,0,m_h^2;\,m_\eta^2,\,m_\Psi^2,\,m_\eta^2),
\qquad
y^h_{\mu e,R} = 0,
\end{equation}
and $C_0$ is the standard scalar three--point Passarino--Veltman function~\cite{tHooft:1978jhc},
\begin{equation}
C_0(0,0,m_h^2;\,m_\eta^2,\,m_\Psi^2,\,m_\eta^2)
= \int_0^1\!\!dx\!\!\int_0^{1-x}\!\!dy\,
\frac{1}{m_\eta^2 + y(m_\Psi^2 - m_\eta^2) - x y\,m_h^2}.
\label{eq:C0param}
\end{equation}

Integrating out the Higgs field at tree level generates the effective four--fermion operator
\begin{equation}
\mathcal{L}_{\text{eff}} \supset 
- \sum_q \frac{y^h_{\mu e,L}\,y_q^h}{m_h^2}\,(\bar e P_L \mu)(\bar q q)
+ \text{h.c.},
\end{equation}
from which the corresponding Wilson coefficients as defined in Eq.~(1) of Ref.~\cite{Kitano:2002mt} are identified as
\begin{equation}
g_{LS}^{(q)} = -\,\frac{\sqrt{2}}{G_F}\,
\frac{y_q^h\,y^h_{\mu e,L}}{m_h^2},
\qquad
g_{RS}^{(q)} = 0.
\label{eq:gLSq}
\end{equation}

\bibliographystyle{JHEP} 
\bibliography{Refs}

\end{document}